# Magnetic Properties of Galactic Molecular Clouds

Philip C. Myers[1], Ian W. Stephens[2,1], Simon Coudé[3,2,1], Nicole Karnath[4,1], & Howard A. Smith[1]

1. Center for Astrophysics | Harvard and Smithsonian, 60 Garden Street, Cambridge MA 02138 USA; pmyers@cfa.harvard.edu
2. Department of Earth, Environment, and Physics, Worcester State University, Worcester, MA 01602, USA
3. National Observatory of Japan, National Institute of Natural Sciences, 2-21-1 Osawa, Mitaka, Tokyo 181-8588, Japan
4. Space Science Institute, 4765 Walnut St., Ste. B., Boulder, CO 80301, USA

## ABSTRACT

The earliest studies of molecular clouds (MCs) showed scaling relations between mass and radius $M \propto R^s$ with exponent $s \approx 2$, known as the third Larson relation. More recent CO surveys indicate $2 \lesssim s \lesssim 3$ in the inner Milky Way (MW) and nearby galaxies. These observations appear consistent with the virial equilibrium (VE) of magnetically subcritical clouds, bound by the pressure $P_e$ of the midplane ISM. The VE equations for such clouds are consistent with $2 < s < 3$ for magnetically subcritical clouds, but not for magnetically critical, supercritical, or nonmagnetic clouds. We fit the subcritical VE equations to $M - R$ trends observed in CO surveys of ~1500 clouds, with correlation coefficients 0.8 - 1. The fits indicate median mass-to-flux ratio $\lambda \approx 0.7$, magnetic field strength $\approx 60$ μG, and $P_e \approx 5 \times 10^5$ cm$^{-3}$ K in the inner MW, with lesser values in M33 and M31. For 18 MW "bones," VE analysis assuming likely values of $P_e$ also indicates subcritical values of $\lambda$, as found from polarization observations. These results indicate that many giant molecular clouds (GMCs) are consistent with pressure-bound magnetically subcritical virial equilibrium. If so, GMCs may have low star formation efficiency ($SFE$) because their relatively strong fields cause filaments, clumps and cores to form slowly, while feedback and turbulence disperse them rapidly. The fields in their clumps and cores are likely weak enough to allow star-forming collapse, according to Zeeman and DCF observations, and according to MHD simulations which predict $SFE \approx 0.5 - 2\%$.

## 1. Introduction

Scaling laws have long been discussed to understand the properties of star-forming molecular clouds. Empirical power-law relations among cloud size, velocity dispersion and mass were first identified by Larson (1981; L81). They have been interpreted in terms of observational limitations, equilibrium, gravitational collapse, and turbulence (L81; Solomon et al. (1987) (S87); Elmegreen 1989 (E89); Lombardi et al. 2010; Ballesteros-Paredes et al. 2011; Kritsuk et al. 2013).

The "third Larson relation" between cloud mass $M$ and radius $R$, $M \propto R^s,\ s \approx 2$ was first interpreted by assuming that turbulence drives cloud velocity dispersions to follow a power law similar to $\sigma_{\rm v} \propto R^{1/2}$, and that the clouds are in simple virial equilibrium (L81). Simple virial equilibrium (SVE) neglects magnetic forces and external pressure (Field et al. 2011; F11). These two assumptions predict $M \propto R^s,\ s \approx 2$ where the surface density is constant from cloud to cloud.

Such $M - R$ models based on SVE neglect the pressure that the interstellar medium (ISM) exerts on a cloud (E89; Blitz & Rosolowsky 2004, 2006; Schinnerer & Leroy 2024). Clouds in pressure-bounded virial equilibrium may follow $M \propto R^s,\ s \approx 2$ if they have similar external pressure (Chièze 1987; McKee et al. 1993 (M93)). This origin of $s \approx 2$ has been questioned because the ISM pressure declines significantly with galactic radius $R_{\rm gal}$ (Heyer et al. 2009; Ostriker & Kim 2022 (OK22)).

In recent years a different $M - R$ scaling law has become apparent. Ten surveys in the MW and nearby galaxies indicate $M \propto R^s$, $s = 2 - 3$, for thousands of MCs. Their $M - R$ trends are similar to each other, but their exponents differ from $s = 2$. Plots of these trends, survey details, and slope statistics are given in Section 2.

Not all surveys show $s = 2 - 3$, and some recent surveys of MW clouds show $s \approx 2$ with relatively high precision (e.g. Chen et al. 2020, Lewis et al. 2022). Nonetheless, the numerous surveys of clouds with $s = 2 - 3$ call for a physical explanation.

RD10 interpreted $s = 2 - 3$ as arising from the fractal structure of turbulent cloud gas (Elmegreen & Falgarone 1996). However the assumption that cloud mass can be used as a proxy for its area (RD10) was criticized by Ballesteros-Paredes et al. 2019 (BP19).

BP19 suggested that $s = 2 - 3$ may arise from superposition of clouds, since the line of sight lies along the galactic plane in MW surveys. This explanation should not apply to inclined galaxies, yet $s = 2 - 3$ in the inclined galaxies M33, M31, the LMC, and NGC 613.

An equilibrium model with a turbulent energy cascade predicts "revised Larson relations" with $s = 11/5 = 2.2$ (Cen 2021). Combining a turbulent energy cascade with all terms of the VE equation would be desirable, but it lies outside the scope of this paper.

This paper addresses $M - R$ observations having $s = 2 - 3$ with the equations of VE for a uniform spherical cloud, including magnetic, gravitational, and kinetic energy due to both internal pressure and external pressure $P_e$ (Spitzer 1968, M93). Such VE analysis including magnetic and pressure terms has not been applied to these $M - R$ observations. Its use here is justified because it predicts clouds whose $M - R$ exponents lie in the range $s = 2 - 3$.

Paper 1 (Myers et al. 2025) analyzed the effect of the MW environment on MCs to better understand their low star formation efficiency ($SFE$). This work extends Paper 1 by analyzing the role of magnetic fields in star-forming MCs in the MW and nearby galaxies.

In this paper, Section 2 presents examples of the observed $M - R$ relation with $s = 2 - 3$. Section 3 describes the VE model of pressurized, magnetized clouds used to analyze the observations. It shows that for likely values of $P_e$, GRS clouds are magnetically subcritical in contrast to outer galaxy clouds, which are magnetically supercritical. Section 4 presents VE equations used for fitting observed $M - R$ relations. It presents derived distributions of cloud mass-to-flux ratio, magnetic field strength, and external pressure in the MW, in M33, and in M31. Section 5 compares mass-to-flux ratios for GRS clouds, Zeeman clumps and DCF cores, and Alfvén Mach numbers for GRS clouds. Section 6 estimates magnetic properties of MW "bones," using VE equations for prolate spheroids. Section 7 discusses limitations, uncertainties, and implications, and Section 8 presents the conclusions. The main conclusions are that many GMCs are strongly magnetized, and that their strong fields help explain their low $SFE$.

## 2. $M - R$ relations with $s = 2 - 3$ in the MW and nearby galaxies

Log $M - \log R$ plots having $s = 2 - 3$ are shown in Figure 1 for MW surveys and in Figure 2 for local galaxies. Nearly all of the available surveys are limited to $^{12}$CO and $^{13}$CO observations. Their survey clouds extend over significant areas of each galaxy. To produce these figures we downloaded the data for each reference and applied a linear fit to each plot.

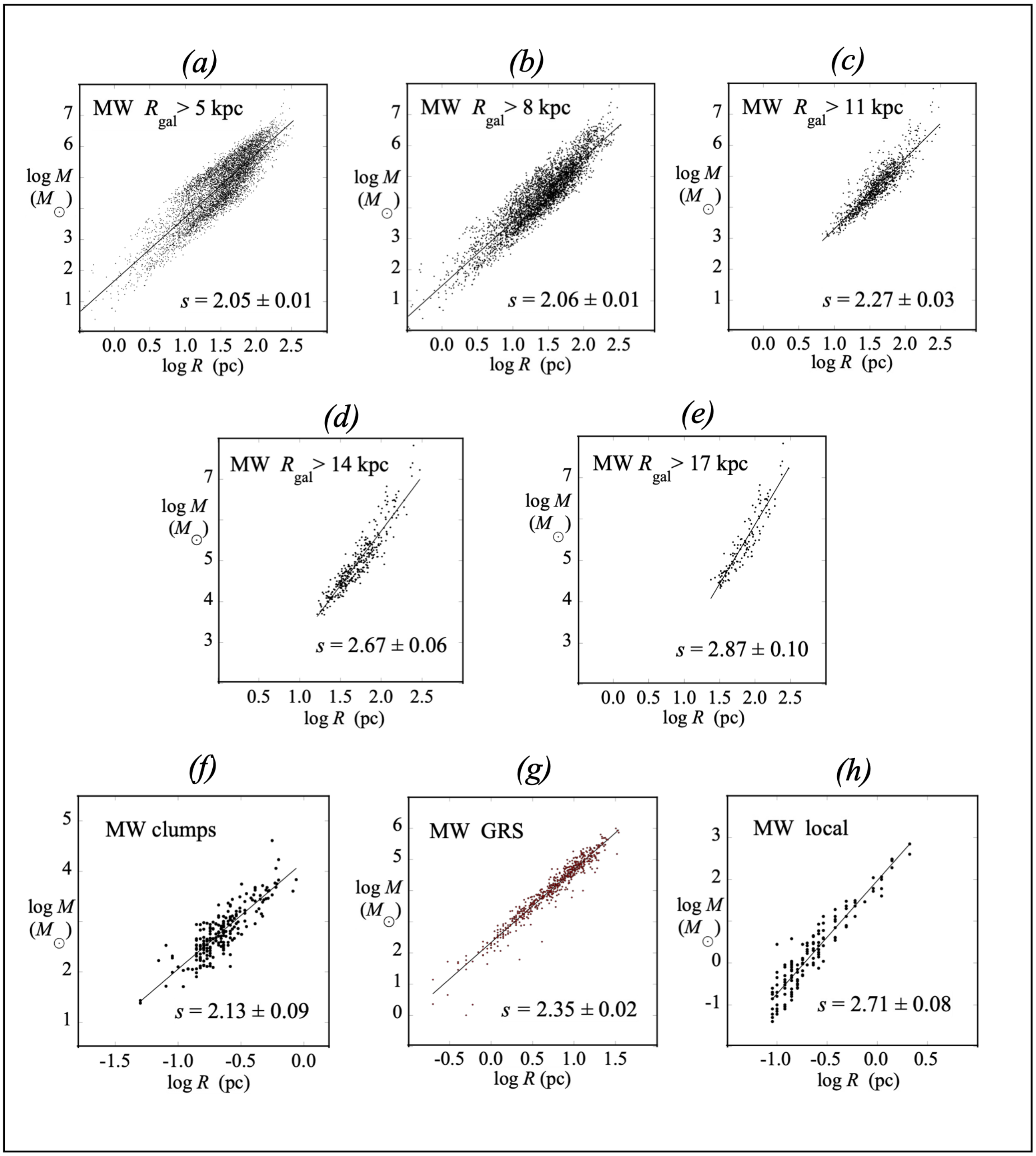

**Figure 1.** Log $M - \log R$ plots for MW clouds having observed slope $s = 2 - 3$, in order of increasing $s$. Plots ($a$) - ($e$) are based on the CO survey data of MD17 and Elia et al. (2025). They cover regions with galactocentric radius $R_{\text{gal,min}} \le R_{\text{gal}} \le 20$ kpc for $R_{\text{gal,min}} = 5,\ 8,\ 11,\ 14,$ and 17 kpc. Plot ($a$) shows two linear trends, as is typical for $R_{\text{gal,min}} \le R_{\text{gal}} \le 20$ kpc when $R_{\text{gal,min}} = 0 - 5$ kpc. Plots ($b$) - ($e$) have $R_{\text{gal,min}} > 5$ kpc and single linear trends. In these plots, $s$ increases with increasing $R_{\mathrm{gal,min}}$, and $s$ spans the range 2 - 3. Plots (*f*) - (*h*) show studies with $s = 2 - 3$ in massive star-forming regions from Traficante et al. (2018), in the GRS (RD10), and in Solar neighborhood clouds (Kainulainen et al. 2011).

Among MW clouds with $s = 2 - 3$, the largest survey has 8107 clouds, with $s = 2.2 \pm 0.2$ based on $^{12}$CO line survey observations (Miville-Deschênes et al. 2017; MD17). In a $^{12}$CO survey of more than 800 clouds in the MW, local arm clouds have $s = 2.18 \pm 0.05$ and Perseus arm clouds have $s = 2.29 \pm 0.03$ (Ma et al. 2021). In the GRS survey $^{12}$CO survey, 583 clouds have $s = 2.35 \pm 0.02$ (RD10). In the Solar neighborhood, 137 clumps have $s = 2.7 \pm 0.1$ based on extinction estimates (Kainulainen et al. 2011). In the galactic plane, 212 massive star-forming clumps have $s = 2.13 \pm 0.09$ based on *Herschel* infrared observations (Traficante et al. 2018).

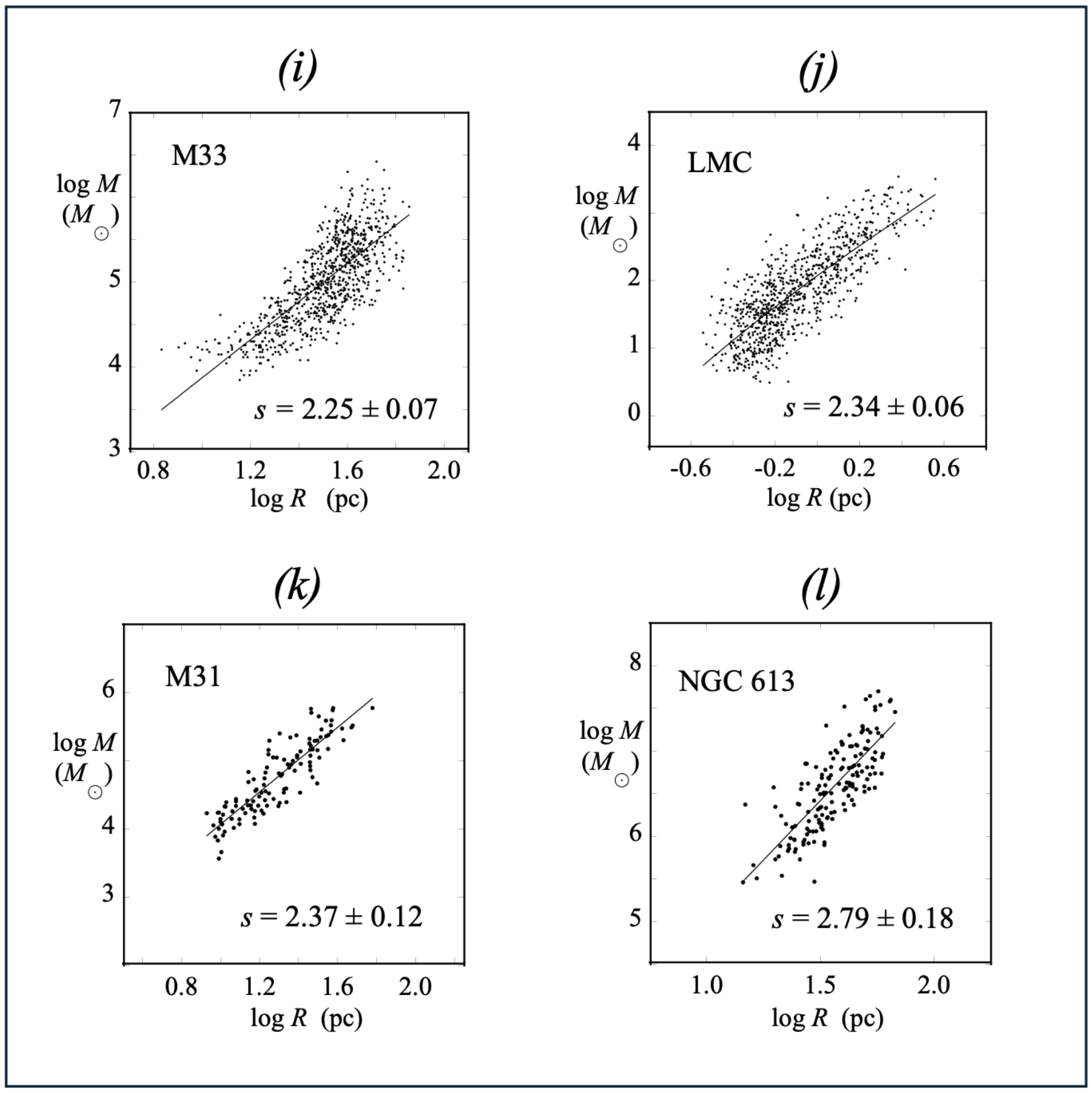

**Figure 2.** Log $M - \log R$ plots of clouds in nearby galaxies having log-log slope $s = 2 - 3$, in order of increasing $s$, for (*i*) M33 (Muraoka et al. 2023), (*j*) the LMC (Wong et al. 2019), (*k*) M31 (Lada et al. 2024), and (*l*) NGC 613 (Choi et al. 2024).

Among nearby galaxy studies with $s = 2 - 3$, CO observations show that 848 clouds in M33 have $s = 2.25 \pm 0.07$ (Muraoka et al. 2023), 839 CO "islands" in the Large Magellanic Cloud have $s = 2.34 \pm 0.06$ (Wong et al. 2019), 117 clouds in M31 have $s = 2.37 \pm 0.12$ (Lada et al. 2024), and 158 clouds in the inner region of NGC 613 have $s = 2.79 \pm 0.18$ (Choi et al. 2024).

Eight of the ten $M - R$ trends described above have very similar slopes $s$, with mean ± standard deviation $2.26 \pm 0.09$ after excluding two outlier trends with $s = 2.7$ and 2.8. This similarity is remarkable since it arises from independent surveys of statistically large samples in multiple galaxies, using different telescopes. The best-defined single trend is from the Galactic Ring Survey (GRS) of 583 $^{13}$CO clouds in the inner Milky Way (MW), with $s = 2.35 \pm 0.02$ (Roman-Duval et al. 2010; RD10).

## 3. $M - R$ equations in magnetic virial equilibrium

This section uses the equations of magnetic PVE (hereafter VE) to analyze the observed $M - R$ relations shown in Section 2. We do not assume that all clouds are necessarily in VE, but we test observations for consistency with VE. VE analysis is especially useful for this paper because it identifies conditions where VE equations predict log $M - R$ slopes $s = 2 - 3$.

We also refer to the virial parameter $\alpha_{\rm vir} = 5\sigma_{\rm v}^2 R/(GM)$ (BM92) but we rely primarily on the VE equation, which gives more specific information on cloud binding when external pressure and mass-to-flux ratio vary. Furthermore, factor-of two observational uncertainties make it difficult to discriminate equilibrium from collapse with values of $\alpha_{\rm vir}$ alone (Chevance et al. 2020; Krumholz et al. 2025).

VE analysis of CO observations is uncertain in part because the CO-mass conversion factor depends on metallicity. In the MW, this dependence may affect estimates of $M$ by up to a factor of ~3 and estimates of $\alpha_{\rm vir}$ by a factor of ~2 (Evans et al. 2021). In the Solar neighborhood, estimates of MC boundedness based on $\alpha_{\rm vir}$ are also biased by up to a factor ~3 due to differing methods of cloud definition and map analysis. This bias can be reduced by an improved analysis procedure (Singh et al. 2021).

Despite such observational uncertainties, the $M - R$ trends in Figures 1 and 2 have well-defined slopes, they apply to thousands of MCs in nearby galaxies, and their VE model fits have statistical confidence, as in Figure 6. Evidently these trends remain significant despite some forms of data uncertainty, as noted in Section 4.2.

In this section, Section 3.1 shows that GRS clouds are consistent with VE for a wide range of external pressure $P_{\mathrm{e}}$. Section 3.2 estimates the normalized mass to flux ratio $\lambda$ of GRS clouds as a function of $P_{\mathrm{e}}$. It shows that variation of $P_{\mathrm{e}}$ also affects the slope and statistical significance of the line width - size relation. Section 3.3 derives a VE expression for $M \propto R^s$ for magnetically subcritical clouds, to model the $M - R$ observations in Section 2.

### 3.1. Equations of virial equilibrium

The virial equation for a uniform spherical cloud of mass $M$, radius $R$, density $\rho$, and velocity dispersion $\sigma_{\mathrm{v}}$, threaded by a uniform magnetic field of strength $B$, and surrounded by a medium with locally constant pressure $P_{\mathrm{e}}$ can be written

$$3(P - P_{\mathrm{e}})V - \frac{3}{5} a \frac{GM^2}{R}\left(1 - \frac{\mathcal{M}}{|\mathcal{W}|}\right) = 0 \qquad (1)$$

where $P = \rho\sigma_{\mathrm{v}}^2$ is the uniform internal pressure, $V$ is the cloud volume, $a$ is the cloud structure form factor, $\mathcal{M}$ is the net magnetic energy and $\mathcal{W}$ is the gravitational binding energy (Spitzer 1968, Elmegreen 1989 (E89), Bertoldi & McKee 1992 (BM92), McKee et al. 1993 (M93)). The form factor $a$ quantifies the gravitational energy change due to the internal density structure of the cloud (BM92, MD17). We assume $a = 1$ as for a uniform sphere. This assumption is consistent with most molecular cloud surveys, which report the radius of a uniform spherical model of each observed cloud (e.g. RD10, MD17). Furthermore, for a centrally concentrated sphere with density-radius law $\rho \propto r^{-1}$ the form factor increases only slightly, by a factor of 1.1 (BM92). For highly elongated clouds (Zucker et al. 2018 (Z18), Soler et al. 2021) we use a virial equation to describe a uniform prolate spheroid, in Section 6.

In equation (1) the interstellar medium pressure $P_{\mathrm{e}}$ in the midplane of a disk galaxy is a key factor in the gravitational binding of MCs (E89, BM92, Sun et al. 2020, OK22, Ganguly et al. 2024). The magnetic energy $\mathcal{M}$ is a significant form of support against gravity in molecular clouds (Mestel & Spitzer 1956, Crutcher 2012, Hennebelle & Inutsuka 2019, Seta & McClure-Griffiths 2025).

Equation (1) does not include the virial surface fluid stress term, the surface magnetic stress, and the bulk flow term. These terms can make order-of-unity changes to the sum of terms in equation (1) (Dib et al. 2007). However they are difficult to extract from observations, and they are usually neglected in virial analysis (Krumholz & Federrath 2019, KF19). Also, equation (1) does not include scaling due to a turbulent energy cascade, as discussed in Sections 1 and 7.1.2.

In equation (1) the ratio of magnetic to gravitational energy $\mathcal{M}/|\mathcal{W}|$ is in turn a function of the normalized ratio of mass $M$ to flux $\Phi$, $\lambda \equiv (M/\Phi)/(M/\Phi)_{\mathrm{cr}}$. The critical value $(M/\Phi)_{\mathrm{cr}} = c_{\Phi} G^{-1/2}$ has been estimated using $c_{\Phi} = \zeta\sqrt{5}/(3\pi)$ (Mouschovias & Spitzer 1976). Assuming $\zeta$ = 0.53 for a sphere (Strittmatter 1966, Peters et al. 2011), $c_{\Phi} = 0.13$. This critical mass-to-flux coefficient is similar to other analytic and numerical estimates, $c_{\Phi} = 0.12$ (Tomisaka et al. (1988) and $c_{\Phi} = 1/(2\pi) = 0.159$ (Nakano & Nakamura 1978). With these values of $c_{\Phi}$, $\mathcal{M}/|\mathcal{W}| \approx \lambda^{-2}$.

A cloud is subvirial, virial, or supervirial when the RHS of equation (1) is $< 0$, 0, or $> 0$, corresponding to global gravitational collapse, equilibrium, or expansion. A cloud is "gravitationally bound" if it cannot expand when $P_{\mathrm{e}} = 0$. Then the inward force of gravity exceeds the outward forces of kinetic and net magnetic pressure, tending to collapse, or they are equal, maintaining equilibrium. The cloud is "bound" or "pressure-bound" if it cannot expand when $P_{\mathrm{e}} > 0$, where in this case the inward forces are due to both gravity and external pressure (BM92). Comparing gravity and magnetic pressure, a cloud may be nonmagnetic ($\lambda = \infty$), magnetically supercritical ($1 < \lambda < \infty$), magnetically critical ($\lambda = 1$), or magnetically subcritical ($\lambda < 1$), in order of increasing magnetic importance (Mouschovias & Spitzer 1976).

To illustrate the difference between magnetically supercritical and subcritical VE, equation (1) can be written

$$3M\sigma_{\mathrm{v}}^2 - 4\pi R^3 P_{\mathrm{e}} - \frac{3GM^2}{5R}\mu_{\mathrm{sup}} = 0 \qquad (2)$$

$$3M\sigma_{\rm v}^2 - 4\pi R^3 P_{\rm e} + \frac{3GM^2}{5R}\mu_{\rm sub} = 0 \qquad (3)$$

where the mass-to-flux indices are $\mu_{\rm sup} \equiv 1 - \lambda^{-2}$ for magnetically supercritical clouds and $\mu_{\rm sub} \equiv \lambda^{-2} - 1$ for magnetically subcritical clouds, and where each index is defined to be > 0. The change in sign of the third term between equation (2) and (3) leads to significantly different properties for magnetically supercritical and subcritical solutions for $M(R)$, as discussed in Section 3.3.

We estimate $\lambda$ from observed properties and from the VE model, from equations (2) and (3):

$$\sigma_{\rm v,vir}^2 = \frac{4\pi R_{\rm obs}^3 P_{\rm e}}{3M_{\rm obs}} \pm \frac{GM_{\rm obs}}{5R_{\rm obs}}\mu \quad . \qquad (4)$$

In equation (4) $\sigma_{\rm v,vir}^2$ is the virial value of $\sigma_{\rm v}^2$. $R_{\rm obs}$ and $M_{\rm obs}$ are observed values of $R$ and $M$, and $P_{\rm e}$ is the assumed value of external pressure. The + sign indicates $\mu = \mu_{\rm sup}$ and the - sign indicates $\mu = \mu_{\rm sub}$.

If a VE solution exists, our solution procedure finds the median value of $\lambda$ for which the observed and predicted values of $\sigma_{\rm v}^2$ are equal. For each chosen value of $P_{\rm e}$, we find $\lambda$ by trial and error (TE), reversing direction after each overshoot, until the sample median of $\left(\sigma_{\rm v,vir}/\sigma_{\rm v,obs}\right)^2$ equals unity. This procedure is applied to the 583 clouds of the GRS survey (RD10). The GRS sample is selected because it describes MCs in the inner MW from $R_{\rm gal} = 3 - 8$ kpc, and because its $\log M - \log R$ slope has the smallest fitting error among all those in Figures 1 and 2, $\sigma_s = 0.02$. The fiducial pressure is chosen to be representative of pressures in the GRS range of $R_{\rm gal}$, $P_{\rm e,0}/k = 3.57 \times 10^5$ cm$^{-3}$ K at $R_{\rm gal} \approx 4$ kpc (OK22). A similar estimate of inner-MW pressure is obtained from equation (12) of Franco et al. (2025). Scaling $P_{\rm e}$ at the Solar neighborhood by the ratio of GRS surface densities at $R_{\rm gal} = 4.5$ kpc and 7.5 kpc (RD10 Figure 6) yields $P_{\rm e}/k = 2.5 \times 10^5$ cm$^{-3}$ K. For completeness we also find VE solutions for $\lambda$ assuming that clouds with GRS values of $M, R,$ and $\sigma_{\rm v}$ are subject to pressures extending above and below those expected at 4 kpc, $P_{\rm e}/P_{\rm e0} \approx 0 - 1.5$.

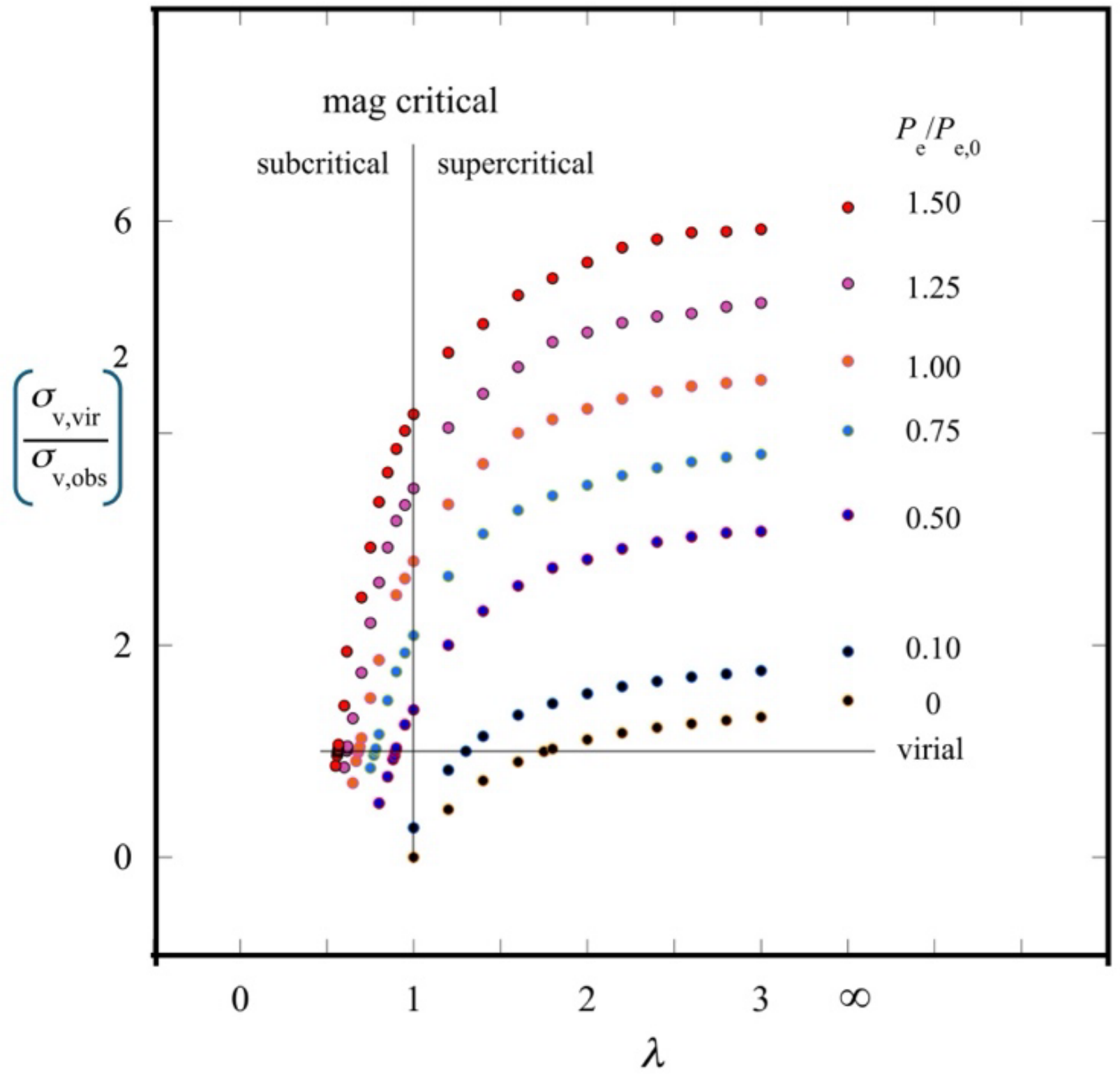

**Figure 3.** Median squared ratio of the virial velocity dispersion to the observed velocity dispersion $(\sigma_{v,vir}/\sigma_{v,obs})^2$ in GRS clouds, as a function of normalized mass to flux ratio $\lambda$ and external pressure $P_e$, according to equation (4). Colored circles indicate the likely range of $P_e$ in the inner MW covered by the GRS, $P_e/P_{e0} = 0.5 - 1.5$, with $P_{e,0}/k = 3.57 \times 10^5$ cm$^{-3}$ K at the galactocentric radius 4 kpc (OK22). Black circles approximate lower-pressure values in the Solar neighborhood and the outer Galaxy ($P_e/P_{e0} \approx 0.1 - 0$). The ratio $(\sigma_{v,vir}/\sigma_{v,obs})^2$ equals unity when the VE model is consistent with the cloud data, i.e. when a circle lies on the "virial" line.

Figure 3 shows that a typical cloud with observed GRS values of $M, R,$ and $\sigma_v$ can be either virial and gravitationally bound, virial and pressure-bound, subvirial or supervirial, depending on its associated values of $P_e$ and $\lambda$.

As $P_e$ increases in Figure 3, the ratio $(\sigma_{v,vir}/\sigma_{v,obs})^2$ for GRS clouds crosses the virial line as $\lambda$ decreases. We first assume $P_e/P_{e0} \approx 0$ to approximate the ISM pressure in the outer MW, since $P_e$ scales with the exponentially declining surface density of molecular gas (Franco et al. 2025). Then Figure 3 shows that a GRS cloud with $P_e/P_{e0} \approx 0$ is in VE when $\lambda \approx 1.75$ : it is gravitationally bound and magnetically supercritical.

As $P_e$ increases to $0 < P_e/P_{e0} \lesssim 0.3$ as in the Solar neighborhood , a GRS cloud is no longer gravitationally bound, but it is virial and pressure-bound if it is magnetically supercritical, with $1 < \lambda < 1.75$. At higher pressures $0.3 < P_e/P_{e0} \lesssim 1.5$ including those in the inner MW, a GRS cloud requires stronger fields to be virial and pressure-bound. Then it must be magnetically subcritical, with $0.5 < \lambda < 1$.

Figure 3 shows that some GRS clouds with high $P_e$ and weak fields are subvirial: all clouds with $P_e/P_{e0} \gtrsim 0.5$ and $\lambda \gtrsim 1.75$ lie above the virial line. They tend to collapse because the inward force of their self-gravity is greater than their outward magnetic and kinetic pressure forces.

In contrast, Figure 3 shows that some GRS clouds with low $P_e$ are supervirial (they lie below the virial line). When these supervirial clouds have $P_e \approx 0$ they are gravitationally unbound and they tend to expand. They resemble clouds identified by Evans et al. (2021) based on virial parameter analysis. In Figure 3 all of the gravitationally unbound clouds are magnetically supercritical (they lie to the right of the magnetically critical line). However if their $P_e$ increased until they were pressure-bound in VE, they would require stronger fields to satisfy VE. Then they would tend to become magnetically subcritical.

In summary, GRS clouds with $P_e \approx 0$ and $\lambda > 1.75$ are subvirial. For virial GRS clouds, those with $P_e \approx 0$ require $\lambda \approx 1.75$ to be gravitationally bound, and those with $0 < P_e/P_{e0} \lesssim 0.3$ require $\lambda \approx 1 - 1.75$ to be bound by gravity and external pressure. For greater and more realistic inner-galaxy pressures, $0.3 < P_e/P_{e0} < 1.5$, GRS clouds require stronger fields to be bound, with magnetically subcritical values $0.5 < \lambda < 1$.

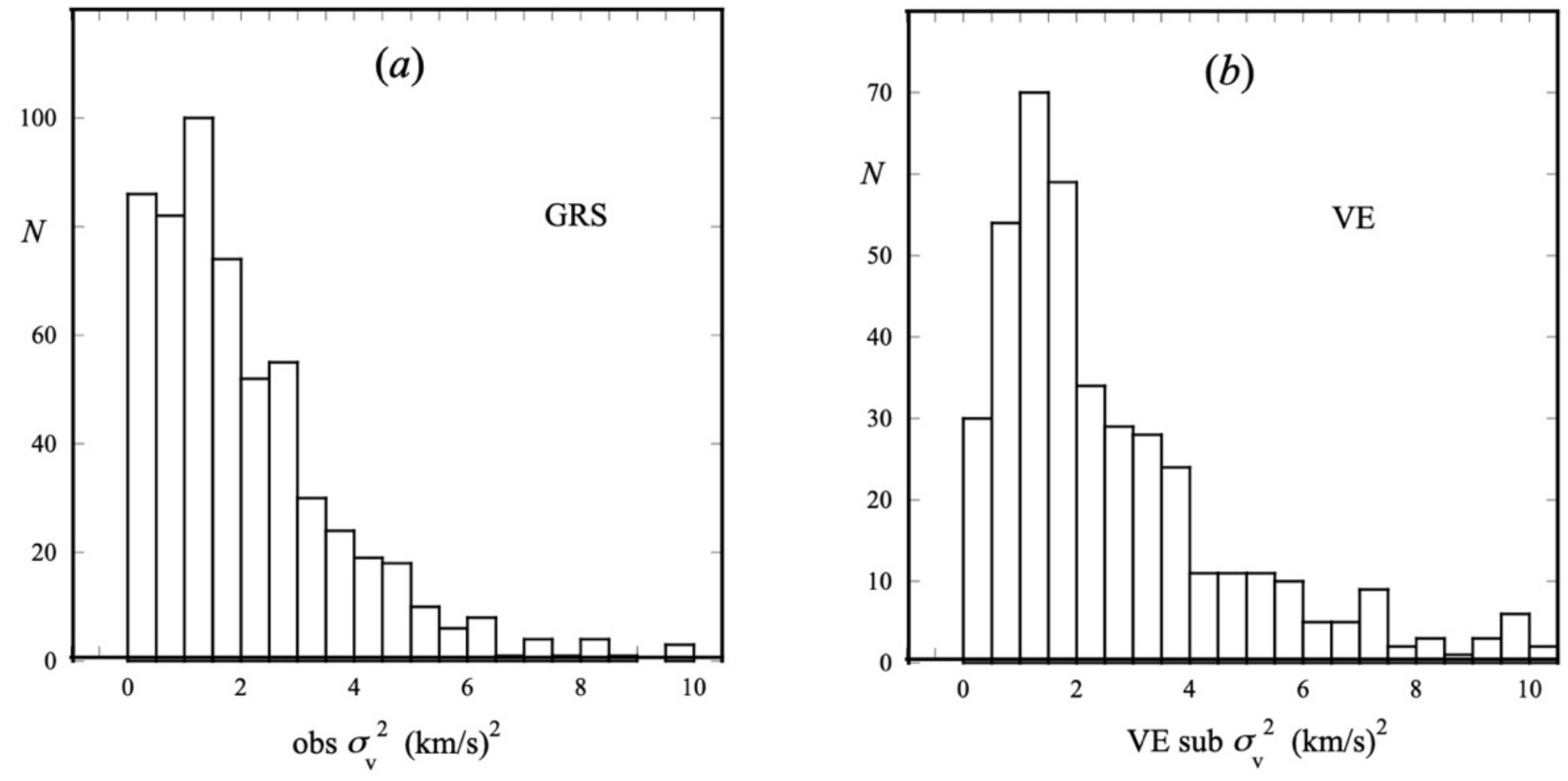

**Figure 4.** Distributions of squared velocity dispersion $\sigma_v^2$ for (*a*) 583 observed GRS clouds (RD10), and (*b*) the VE model solution in Figure 3, with $\lambda = 0.68$ when the external pressure is $P_{e,0}/k = 3.57 \times 10^5$ cm$^{-3}$ K.

The quality of the TE solutions in Figure 3 is indicated by the consistency between the observed and predicted distributions of squared velocity dispersions. This consistency is shown by the similarity of distribution modes and shapes in Figure 4.

### 3.2. Line-width-size relations of virial clouds

The relative values of $P_e$ in equation (4) influence the line-width-size (LWS) relations of virial clouds. When $P_e << P_{e0}$, equation (4) indicates that a self-gravitating cloud follows $\sigma_{v,vir}^2 = \pi G \Sigma R \mu / 5$ where $\Sigma$ is the cloud surface density. Then $\sigma_{v,vir}^2(R)$ approximates the dependence observed for Solar neighborhood and outer galaxy clouds (L81, Sodroski 1991, Brand & Wouterloot 1995, May et al. 1997, Heyer et al. 2001; H01). Virial model estimates match these trends (E89, Heyer et al. 2009, Heyer & Dame 2015).

However, at the galactocentric radius $R_{gal} = 4.5$ kpc where the GRS survey finds the greatest surface density (RD10), $P_e$ is typically greater by an order of magnitude than for clouds in the Solar neighborhood, as noted above. As a result the typical inner-galaxy GRS cloud is magnetically subcritical as in the lower left of Figure 3, while the typical outer-galaxy cloud is magnetically supercritical as in the lower right of Figure 3. In the subcritical case the outward pressure of the cloud magnetic field exceeds the inward pressure due to self-gravity and external pressure by more than in the supercritical case. Then the dependence of $\sigma_{\rm v}^2$ on $R$ may be reduced significantly from supercritical to subcritical clouds.

Figure 5 illustrates these points. In ($a$) the LWS relation observed by the GRS has a slope which is shallow and poorly defined. These properties are approximated by the virial model in ($b$) using GRS values of $M$ and $R$, the assumed value $P_{\rm e} = P_{\rm e0}$, and the subcritical estimate $\lambda = 0.56$ from Figure 3. Then the opposing pressure and self-gravity terms in equation (4) are nearly equal. In contrast, ($c$), a magnetically supercritical virial model from Figure 3 with GRS values of $M$ and $R$, $P_{\rm e} = 0$ and $\lambda = 1.75$ predicts a LWS relation with steep slope and strong correlation.

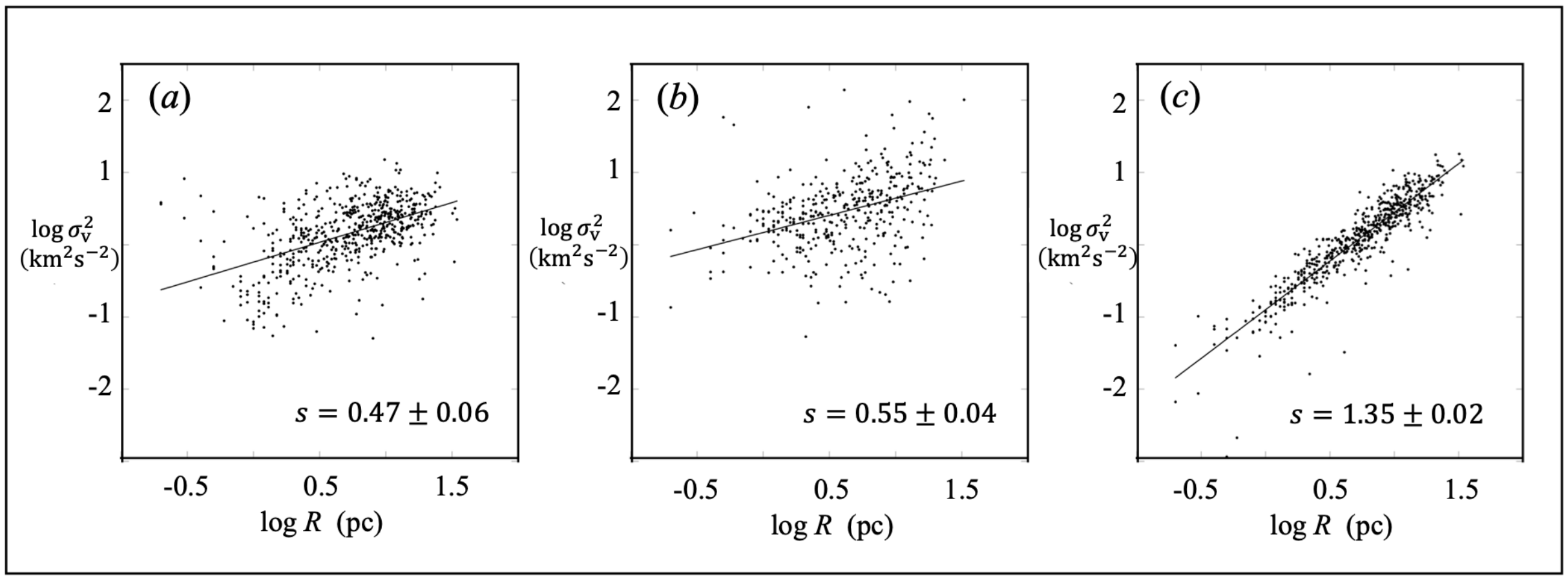

**Figure 5.** Line width-size (LWS) plots associated with changing cloud external pressure $P_{\rm e}$. Panel ($a$) shows the LWS plot $\log \sigma_{\rm v}^2$ *v.* $\log R$ for molecular clouds observed in the inner galaxy (RD10) with a linear fit, having slope 0.6 and correlation coefficient 0.5. Panel ($b$) shows a VE LWS plot predicted by equation (4) using GRS values of $M$, $R$, and $P_{\rm e}$. Panel ($c$) shows the VE LWS relation resulting from equation (4) and GRS values of $M$ and $R$ as in ($b$), but assuming that $P_{\rm e}$ approximates outer galaxy values, with $P_{\rm e}/k \approx 0$ and $\lambda = 1.75$ as in Figure 3. The resulting LWS has steeper slope and greater correlation coefficient than for clouds with lower $P_{\rm e}$ in ($a$) or ($b$).

LWS relations with shallow slope for pressure-confined clouds and steep slope for self-gravitating clouds are also seen in a CO survey of outer-galaxy clouds (Benedettini et al. 2020). These results and those in Figure 5 suggest that LWS relations can be approximated by the VE equations, when their magnetic and pressure terms are included with realistic estimates of $\lambda$ and $P_e$. Then inner galaxy clouds with high $P_e$ are subcritical and have shallow, poorly defined LWS slopes. But outer galaxy clouds with low $P_e$ are supercritical, with steep, well defined LWS slopes.

### 3.3. Virial equilibrium model of clouds with $M \propto R^s$, $s = 2 - 3$

This section describes the $\log M - \log R$ slopes $s_{\rm vir} \equiv d \log M / d \log R$ of VE clouds for comparison with the observed log-log slopes $s = 2 - 3$. Henceforth we use "slope" for "log-log slope."

For VE clouds, equation (4) is quadratic in cloud mass $M$. It can be solved approximately for $M(R)$ when $\sigma_{\rm v}$, $P_{\rm e}$, and $\mu$ have nearly fixed values. Its solution has counterparts to the stable and unstable solutions for the Bonnor-Ebert sphere (Nakano et al. 1978, F11). Here we discuss the VE counterparts to the stable solutions having $M > 0$. For clouds which are nonmagnetic ($\lambda = \infty$) or magnetically supercritical ($1 < \lambda < \infty$), $M(R)$ has two solutions ("+" and "-"). The slope of the "+" solution becomes more negative with increasing $R$ , i.e. $0 \geq s \geq -\infty$. In contrast, the slope of the "-" solution becomes more positive with increasing $R$, $3 \leq s \leq \infty$. For clouds which are magnetically critical ($\lambda = 1$), $M(R)$ has one solution, with constant slope $s = 3$. For clouds which are magnetically subcritical ($0 < \lambda < 1$), $M(R)$ also has one solution, whose slope becomes shallower with increasing $R$, $3 > s > 2$.

Among these four regimes of $\lambda$, only the magnetically subcritical solutions ($0 < \lambda < 1$) match the observed range of slopes $s = 2 - 3$. Thus we select the magnetically subcritical solution to the VE equation (4) to fit observed cloud samples with $M \propto R^s$, $s = 2 - 3$. The origins of slopes $s = 2$ and $2 < s < 3$ are discussed further in Section 7.1.1.

The solution for $M(R)$ to the magnetically subcritical VE equation (3) is

$$M = \frac{5\sigma_{\rm v}^2 R}{2G\mu_{\rm sub}} \left[ \sqrt{1 + \left(R/R_{0,\rm sub}\right)^2} - 1 \right] \quad . \qquad (11)$$

In equation (11) the mass-to-flux index is $\mu_{\rm sub} \equiv \lambda^{-2} - 1$ and the scale length $R_{0,\rm sub}$ is

$$R_{0,\rm sub} \equiv \left(\frac{15}{16\pi P_{\rm e} G \mu_{\rm sub}}\right)^{1/2} \sigma_{\rm v}^2 \quad . \qquad (12)$$

As a check, solving equation (11) for $\sigma_v^2$ recovers the original VE equation (3).

The slope $s_{vir}$ of the subcritical virial $M - R$ relation is obtained by differentiating the log of equation (11) with respect to $\log R$. Then

$$M = \frac{5\sigma_v^2}{2G\mu_{sub}} R^{s_{vir}} \tag{13}$$

where

$$s_{vir} \equiv 2 + \left[1 + \left(R/R_{0,sub}\right)^2\right]^{-1/2} \quad . \tag{14}$$

Thus $s_{vir}$ decreases with increasing $R$ from $s_{vir} = 3$ when $R/R_{0,sub} \ll 1$,

$$M \approx \left[5\sigma_v^2/\left(4GR_{0,sub}^2\right)\right]R^3 \ , \tag{15}$$

to $s_{vir} = 2$ when $R/R_{0,sub} \gg 1$,

$$M \approx \left[5\sigma_v^2/\left(2GR_{0,sub}\right)\right]R^2 \quad . \tag{16}$$

If a sample of clouds has a sufficiently wide range of radii $R$ below and above their typical scale length $R_{0,sub}$, $M \propto R^3$ for the smallest clouds and $M \propto R^2$ for the largest clouds. This property offers a simple explanation for the prevalence of cloud samples having $s = 2 - 3$, described in Section 2.

**3.4. Virial parameter for magnetically subcritical clouds** The virial parameter $\alpha_{vir} = 5\sigma_v^2 R/(GM)$ (BM92) can be estimated in terms of $s_{vir}$ for a magnetically subcritical cloud. Combining equations (11) and (14) with the definition of $\alpha_{vir}$ gives

$$\alpha_{vir} = \frac{2\mu_{sub}}{(s_{vir}-2)^{-1}-1} \quad . \tag{17}$$

In a sample of clouds fit by a linear $\log M - \log R$ slope $s$ as in Figures 1 and 2, a cloud whose VE model slope $s_{\mathrm{vir}}$ matches $s$ may be considered typical of the cloud sample. Then substituting $s$ for $s_{\mathrm{vir}}$ in equation (17) gives the virial parameter typical of the sample, $\alpha_{\mathrm{vir,typ}} = 2\mu_{\mathrm{sub}}/[(s-2)^{-1} - 1]$. This metric is estimated for three galactic samples in Section 4.4.

## 4. Fitting magnetically subcritical VE equations to mass-radius trends

This section uses the magnetically subcritical VE equations (11) - (16) to fit the $M - R$ trends shown in Section 2. These fits are used to estimate the normalized mass to flux ratio $\lambda$, the magnetic field strength $B_{\mathrm{tot}}$, and the cloud external pressure $P_{\mathrm{e}}$. Section 4.1 presents the fitting equations in log-log form. Section 4.2 illustrates a VE fit to the $M - R$ data for the GRS Survey in Figure 1(*g*). Section 4.3 compares $\lambda$ obtained from the TE solutions in Section 3 and the VE fitting in Section 4. Section 4.4 presents distributions of $\lambda$, $B_{\mathrm{tot}}$, and $P_{\mathrm{e}}$ derived from VE fits to $M - R$ trends in the MW, in M33, and M31.

### 4.1. $\log M - \log R$ fitting equations

The $M - R$ fitting procedure renders equations (11) and (12) in log-log form,

$$\log M = m_1 + \log R + \log\left[-1 + \sqrt{1 + (10^{\log R})^2 (m_2)^{-2}}\right] \qquad (18)$$

where $M$ and $R$ have units $M_{\odot}$ and pc, respectively. The fit and model parameters are related by

$$m_1 = \log\left[\frac{5\sigma_{\mathrm{v}}^2}{2G\mu_{\mathrm{sub}}} \times \frac{(\mathrm{pc})}{(M_{\odot})}\right] \qquad (19)$$

$$m_2 = \left(\frac{15}{16\pi P_{\mathrm{e}} G \mu_{\mathrm{sub}}}\right)^{1/2} \sigma_{\mathrm{v}}^2 \times (\mathrm{pc})^{-1} \qquad (20)$$

where $(M_{\odot}) \equiv 1.99 \times 10^{33}$ g, $(\mathrm{pc}) \equiv 3.09 \times 10^{18}$ cm, with $\sigma_{\mathrm{v}}$, $P_{\mathrm{e}}$, and $G$ in cgs units. Then

$$\lambda = \left[1 + \frac{5\sigma_{\mathrm{v}}^2}{2G 10^{m_1}} \times \frac{(\mathrm{pc})}{(M_{\odot})}\right]^{-1/2}, \qquad (21)$$

$$\lambda = \left[1 + 582 \times 10^{-m_1} \left(\frac{\sigma_{\mathrm{v}}}{\mathrm{km\ s^{-1}}}\right)^{2}\right]^{-1/2}, \qquad (22)$$

$$P_{\mathrm{e}} = \left(\frac{3}{8\pi}\right)\left(\frac{10^{m_1}}{m_2^2}\right)\sigma_{\mathrm{v}}^2 \times \frac{(M_{\odot})}{(\mathrm{pc})^3} \ , \qquad (23)$$

$$P_{\mathrm{e}}/k = (5.83 \times 10^{-3})\left(\frac{10^{m_1}}{m_2^2}\right)\left(\frac{\sigma_{\mathrm{v}}}{\mathrm{km\ s^{-1}}}\right)^2 \times 10^5\ \mathrm{cm^{-3}\ K}\ , \qquad (24)$$

and

$$B_{\mathrm{tot}} = \frac{0.32}{\lambda}\left(\frac{\Sigma}{M_{\odot}\mathrm{pc^{-2}}}\right)\left(\frac{c_{\Phi}}{0.17}\right)^{-1} \mu G \ , \qquad (25)$$

based on $B_{\mathrm{tot}} = \Sigma G^{1/2}(\lambda c_{\Phi})^{-1}$ (M93).

When a $M - R$ trend in a cloud sample is fit by this magnetically subcritical model, equations (22) and (25) predict the magnetic field strength $B_{\mathrm{tot}}$ in terms of the observed values of $M, R,$ and $\sigma_{\mathrm{v}}$, the value of the $M - R$ fit parameter $m_1$, and the mass-to-flux coefficient $c_{\Phi}$. A $M - R$ fit is shown for MW GRS clouds in Figure 6, and their resulting field strength distributions are shown in Figure 8 for clouds in the MW, in M31, and M33.

**4.2. Linear and VE fitting for GRS data**

Figure 6 shows both a linear fit (*a*) and a VE fit (*b*) to the GRS data in Figure 1(*g*). These fits are only slightly different in appearance and fit quality, with VE fit parameters $m_1 = 2.80 \pm 0.06$, $m_2 = 1.19 \pm 0.12$, and with the same correlation coefficient 0.97. The VE fit line shows a slight curvature as the cloud radius increases from $R \ll R_{0,\mathrm{sub}} = 1.19$ pc, where $s_{\mathrm{vir}} = 3.00$, to $R \gg R_{0,\mathrm{sub}}$, where $s_{\mathrm{vir}} = 2.14$. The distributions of $\lambda$, $B_{\mathrm{tot}}$, and $P_{\mathrm{e}}/k$ derived from these fits are shown with their medians and standard deviations in Figure 8.

The relative uncertainty due to statistical fitting error alone is about 4% in $\lambda$ and in $B_{\mathrm{tot}}$, due to fitting uncertainty in $m_1$. Thus $\lambda = 0.62 \pm 0.03$ from equation (22) at the median observed value of $\sigma_{\mathrm{v}}$, and $B_{\mathrm{tot}} = 70 \pm 3\ \mu G$ from equation (25) at the median observed value of $\Sigma$. The relative uncertainty in the median $\sigma_{\mathrm{v}}$ and in the median $\Sigma$ are each about 2% (RD10 Figure 3). The relative uncertainty in $P_{\mathrm{e}}$ due to fitting error is about 25%, due mainly to uncertainty in $m_2$, i.e. $P_{\mathrm{e}}/k = 4.5 \pm 1.1 \times 10^5\ \mathrm{cm^{-3}}$ K at the median observed $\sigma_{\mathrm{v}}$.

If observed cloud mass estimates $M$ have a systematic error of a factor of 2 but no systematic error in radius $R$, the resulting slope $s$ would not change. However the value of $\lambda$ inferred from equation (13) would have a systematic error. Assuming that the typical velocity dispersion has the typical GRS median value $\sigma_{\mathrm{v}} = 1.32\ \ \mathrm{km\ s^{-1}}$ (RD10), equation (13) indicates that a factor of 2 increase in $M$ would increase $\lambda$ from 0.62 to 0.75, while a factor of 2 decrease in $M$ would reduce $\lambda$ from 0.62 to 0.49. This systematic uncertainty would be similar to the standard deviation of the distribution of VE estimates of $\lambda$ in Figure 8(*a*).

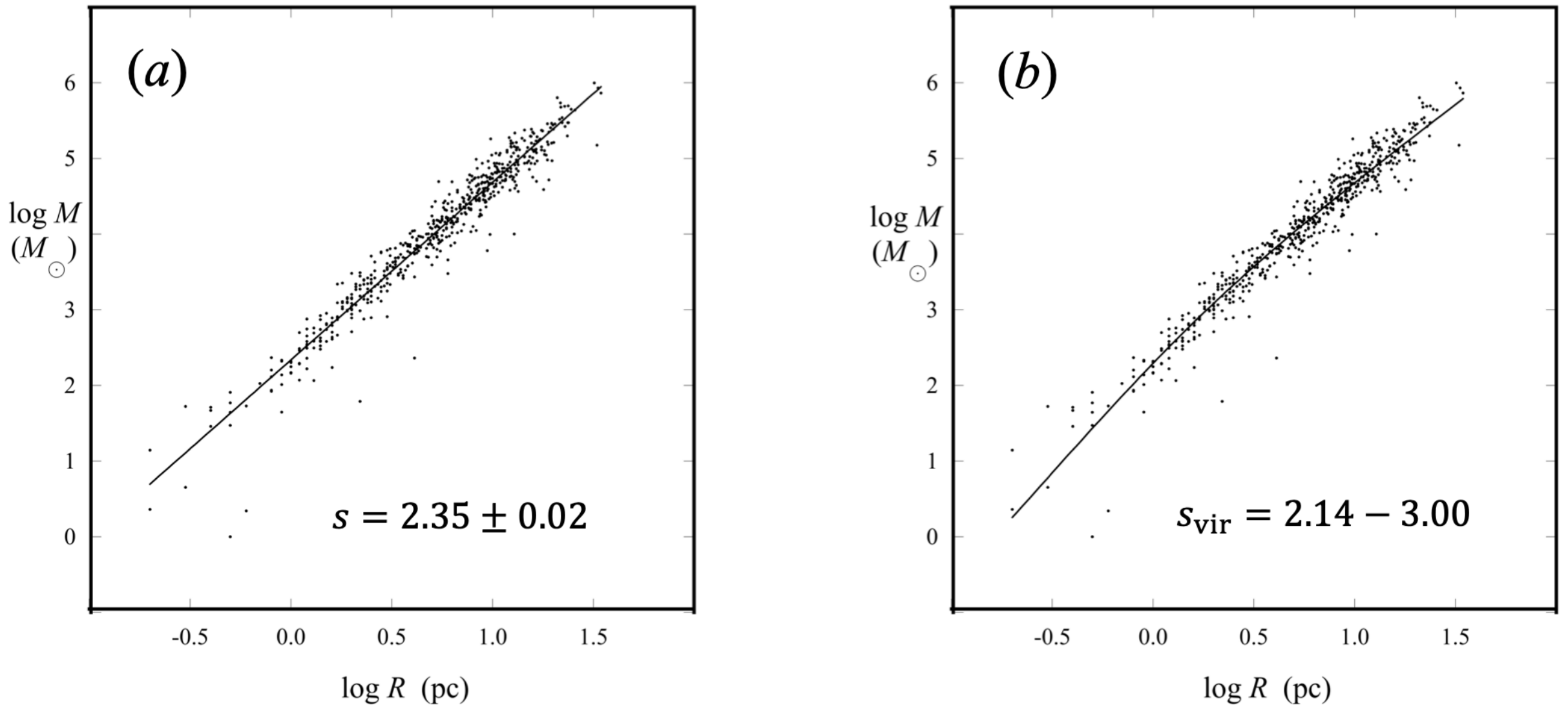

**Figure 6.** Linear fit (*a*) and VE fit (*b*) to the $M - R$ trend in Figure 1(*g*) from the GRS survey with 583 clouds (RD10). The two fits have correlation coefficient 0.97. The data are equally well fit by the linear fit slope $s = 2.35$ and by the range of VE fit slopes $s_{\rm vir} = 2.14 - 3.00$.

### 4.3. Comparison of virial estimates of $\lambda$

The quadratic subcritical VE solutions for $\lambda$ in this section are similar to the trial-and-error (TE) VE solutions for $\lambda$ in Section 3. For the five values of $P_{\rm e}/P_{\rm e0} = 0.5 - 1.5$ assumed in Figure 3, the TE values have mean $\lambda$ = 0.67 and the quadratic subcritical values have mean $\lambda$ =0.71. Their paired values have rms deviation 0.058, with rms deviation/mean ≈ 9%. Thus Figure 7 shows that the VE fit is slightly less accurate than the TE solution. Figure 7 also shows that the VE subcritical fit (open circles) is biased toward lower $\lambda$ when $P_{\rm e}/P_{\rm e0} \lesssim 0.5$. The supercritical TE solutions apply only for low values of $P_{\rm e}/P_{\rm e0}$, as was already seen in the analysis of Figure 3.

The fit of the subcritical VE model to the GRS data in Figure 6(*b*) yields the $P_{\rm e}/k$ distribution in Figure 8(*c*), with median value $P_{\rm e}/k = 4.3 \times 10^5$ cm$^{-3}$ K or $P_{\rm e}/P_{\rm e0} \approx 1.2$. Then the filled and open circles in Figure 7 nearly coincide, thus the VE subcritical fit approximates the TE solution. The error bars show the standard deviation of the VE fit distribution of $\lambda$ in Figure 8($a$). The median value $\lambda = 0.67$ lies about 2-sigma below the critical line $\lambda = 1$, thus the subcritical property of the typical GRS cloud is statistically significant.

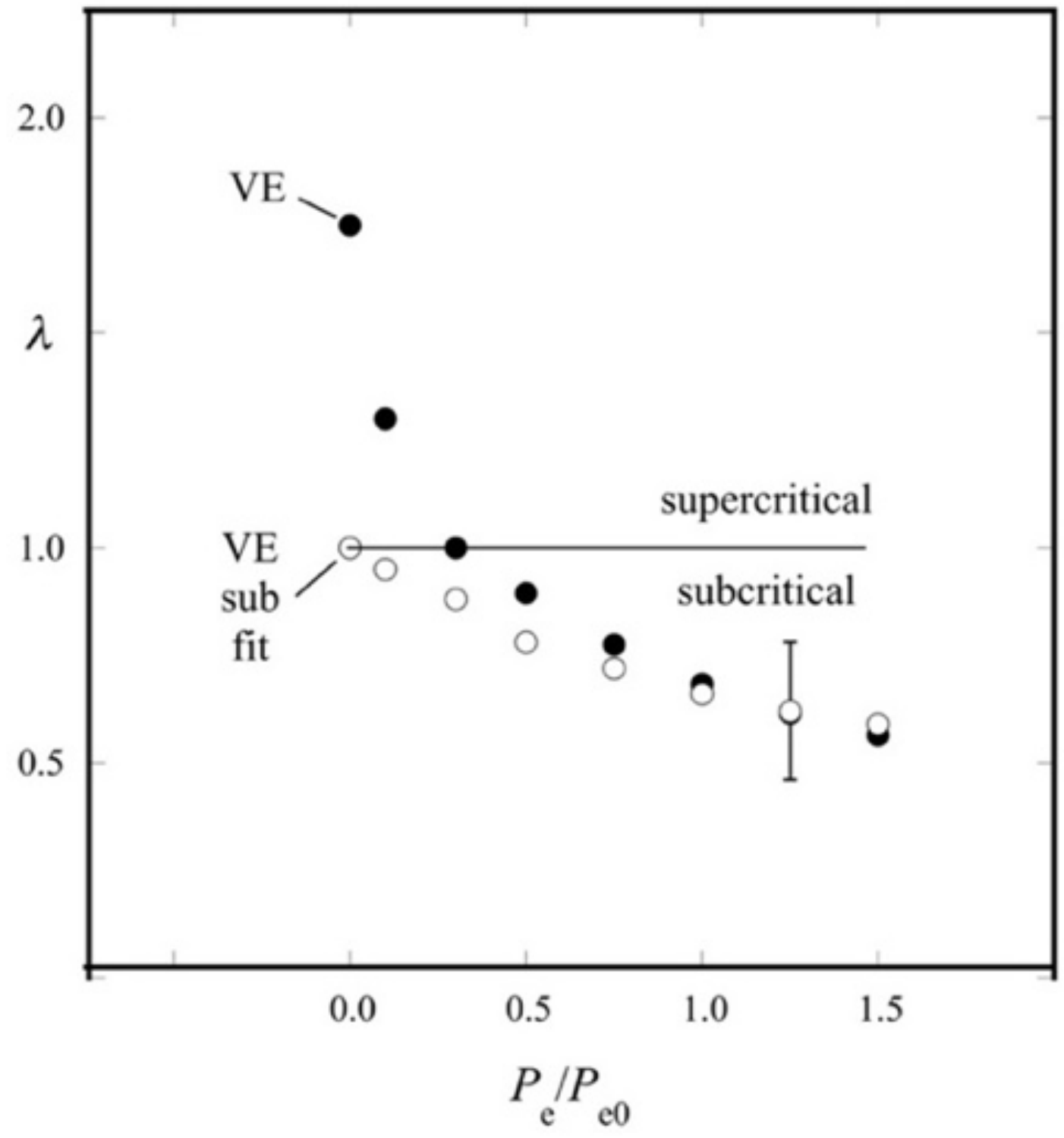

**Figure 7.** Estimates of mass-to-flux ratio $\lambda$ of GRS clouds vs. $P_e/P_{e0}$, the ratio of external pressure to its fiducial value $P_{e0}/k = 5.4 \times 10^5$ cm$^{-3}$ K. The filled circles ("VE") are estimates of $\lambda$ from the TE solutions to the VE equations in Section 3. The open circles ("VE sub fit") are estimates of $\lambda$ from the fits of the magnetically subcritical VE equations to the $M - R$ data in Figure 6($b$). The error bar is based on the fit in Figure 6($b$) and on the distribution in Figure 8($a$).

### 4.4. Distributions of $\lambda$, $B_{tot}$, and $P_e/k$ in the MW, M31, and M33

This section presents distributions of magnetic and external pressure properties derived from VE subcritical model fits to well-defined $M - R$ trends in the MW, M31, and M33, using equations (18)-(25) and the observed values of $M, R$, and $\sigma_v$ from each data sample.

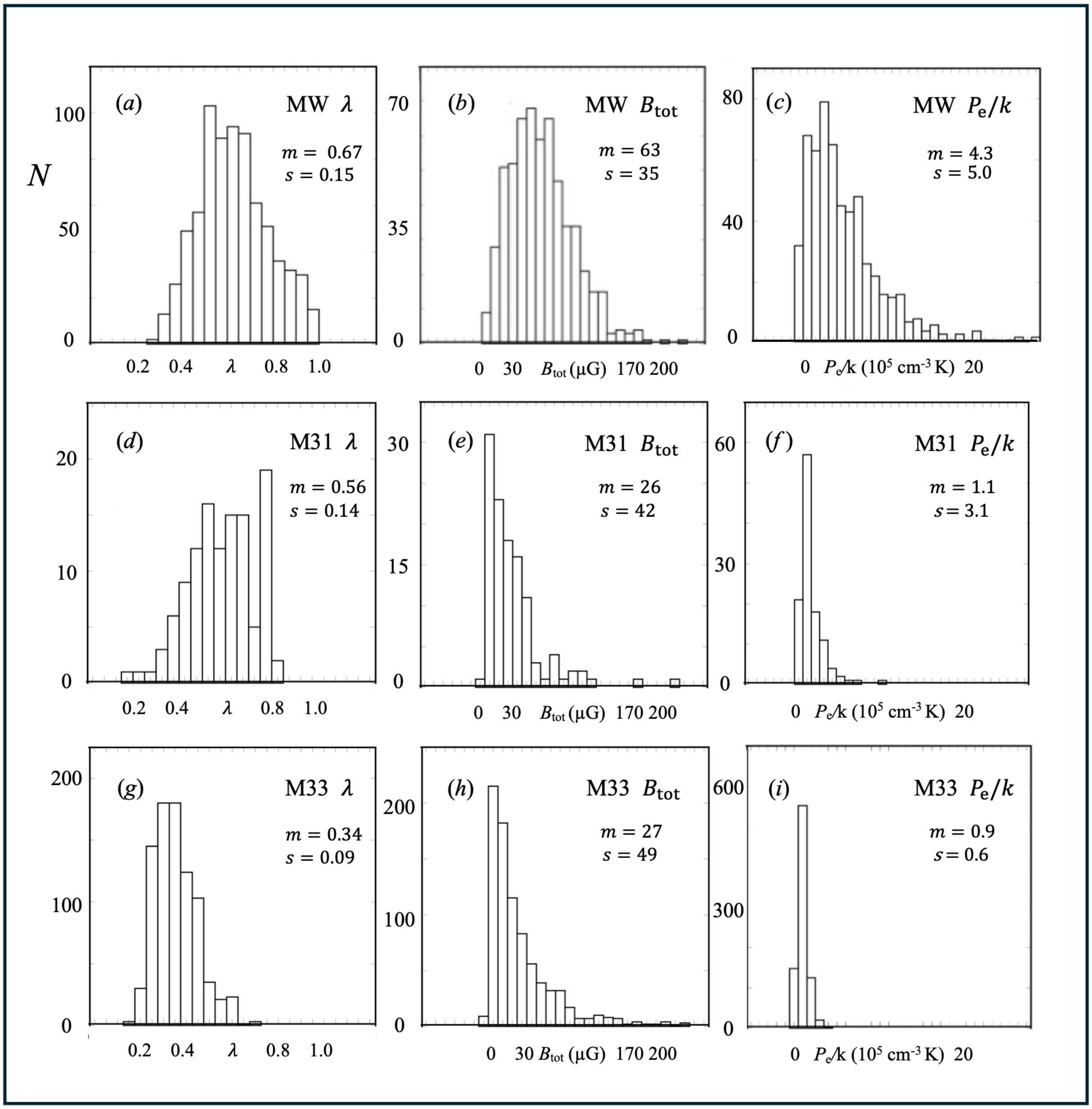

**Figure 8.** Distributions of the number of clouds $N$ in each bin of the mass-to-flux ratio $\lambda$, total magnetic field strength $B_{\rm tot}$, and external pressure $P_{\rm e}/k$ for 583 clouds in the MW (*a*), (*b*), and (*c*); 117 clouds in M31 (*d*), (*e*), and (*f*); and 848 clouds in M33 (*g*), (*h*), and (*i*). Each distribution is based on the magnetically subcritical VE equations (11) and (12) for $M(R)$, and on the fitting equations (18)-(25) for $\lambda$, $B_{\rm tot}$, and $P_{\rm e}/k$. They are applied to the observed data for $M(R)$ in Figure 1(*g*) from RD10, Figure 2(*k*) from L24, and Figure 2(*i*) from M23. Each panel shows the median

$m$ and standard deviation $s$ of the panel distribution, with $B_{\rm tot}$ in units of μG and $P_{\rm e}/k$ in units of $10^5$ cm$^{-3}$ K.

In Figure 8, each distribution of $\lambda$ appears consistent with magnetically subcritical clouds in VE. From galaxy to galaxy, the distributions of $\lambda$, $B_{\rm tot}$, and $P_{\rm e}/k$ have significant overlap, but the distributions become narrower and their typical values decrease from the MW to M31, and from M31 to M33.

This ranking of median $\lambda$, $B_{\rm tot}$, and $P_{\rm e}/k$ from galaxy to galaxy is similar to the ranking of their total molecular cloud mass (MW, 1 ×$10^9$ $M_{\odot}$, Heyer & Dame 2015; M31, 3×$10^8$ $M_{\odot}$, Nieten et al. 2006; and M33, 2×$10^8$ $M_{\odot}$, Gratier et al. 2010). The typical virial parameter defined following equation (17) varies from $\alpha_{\rm vir,typ}$ = 1.3 to 2.6 to 5.1, also indicating a decrease in star formation activity from the MW to M31 to M33. Most distributions have similar asymmetry, with a steeper rise at low values and a shallower tail at high values.

## 5. Magnetic energy metrics

This section compares magnetic energy with gravitational energy through the mass-to-flux ratio $\lambda$ in CO clouds, Zeeman sources, and dense cores in Section 5.1, and magnetic energy with turbulent energy in CO clouds through the Alfvén Mach number $M_A$ in Section 5.2.

### 5.1. Subcritical CO clouds, supercritical Zeeman sources, and supercritical dense cores

Sections 3 and 4 show that many CO survey clouds are magnetically subcritical according to VE analysis. In contrast, most observed molecular cloud Zeeman survey sources are magnetically supercritical (Crutcher 2012; C12), as are most observed low-mass dense cores according to DCF analysis (Myers & Basu 2021, (MB21), Liu et al. 2022 (L22)). The L22 study is a compilation of all 288 previously published DCF estimates, including those of MB21. To simplify nomenclature, henceforth we call CO survey sources "GMCs," molecular cloud Zeeman sources "clumps" and low-mass dense cores "cores."

Although GMCs, clumps and cores are all considered "star-forming," their magnetic criticality applies to very different scales: the typical cloud is much larger, less dense, and more massive than the typical clump or core.

To illustrate this difference in scales, GMCs in the GRS $^{13}$CO cloud survey have median radius $R = 6.1$ pc, density $n = 3 \times 10^2$ cm$^{-3}$, and mass $M = 15{,}100\ M_\odot$ (RD10). In $^{12}$CO observations the typical GMC is also larger, less dense, and more massive, i.e. $R = 25$ pc, $n = 10$ cm$^{-3}$, and mass $M = 38{,}000\ M_\odot$ (MD17). For Zeeman clumps, typical values are $R = 0.2$ pc, $n = 4 \times 10^4$ cm$^{-3}$, and $M = 190\ M_\odot$ (Troland & Crutcher 2008, Crutcher 1999, Falgarone et al. 2008). For DCF cores, typical values are $R = 0.1$ pc, $n = 3 \times 10^5$ cm$^{-3}$, and $M = 40\ M_\odot$ (L22). Thus the typical GRS CO cloud is larger than a typical clump or core by a factor ~100-300, less dense by a factor ~100-1000, and more massive by a factor ~80-400.

Given these differences in size scale and magnetic criticality, it is important for our understanding of star formation efficiency ($SFE$) in GMCs to know the criticality of clumps and cores embedded in GMCs. Such dense regions are prevalent in GMCs. The GRS clouds have numerous embedded clumps and cores (Rathborne et al. 2009), as do ATLASGAL survey clouds (Wienen et al. 2012), and infrared dark clouds in the ASHES survey (Morii et al. 2023).

However, systematic estimates of $\lambda$ in GMCs and their clumps and cores are not available, due to observational limitations of sensitivity, resolution, and line-of-sight confusion (Pattle et al. 2023; P23). Instead we use literature estimates from OH Zeeman observations (C12), and DCF estimates based on near-infrared and far-infrared polarization in Solar neighborhood clumps and dense cores (L22, MB21). We assume that their magnetic properties are representative of the condensations embedded in GRS survey clouds.

Figure 9 shows 583 magnetic field strengths $B_{\rm pos}$ derived from VE analysis of GMCs in the GRS (RD10), 37 estimates of $B_{\rm los}$ from OH Zeeman observations of clumps (C12), and 288 estimates of $B_{\rm pos}$ from DCF analysis of far infrared polarization, primarily from cores (L22). The field strengths are plotted in log-log form against the associated column density $N$ in Figure 9($a$) and against the associated number density $n$ in Figure 9($b$).

Figure 9($a$) shows that when DCF cores and Zeeman clumps have column densities greater than GMC column densities ($\log N \gtrsim 22.3$), their field strengths exceed the GMC field strengths, but they tend to lie below the magnetically critical line ($\lambda = 1$) while the GMC field strengths lie above it. Thus these cores and clumps are magnetically supercritical as found by C12 and L22, yet their fields tend to be stronger than in the magnetically subcritical GMCs.

Figure 9($b$) shows that most DCF cores and Zeeman clumps have much greater number densities than do GMCs, as described above, and also that their observed field strengths are typically

greater than the VT estimates for GMCs, as expected from their relative column densities in Figure 9($a$). The least-squares fit to the scaling of field strength with density has exponent 0.5, similar to that in L22.

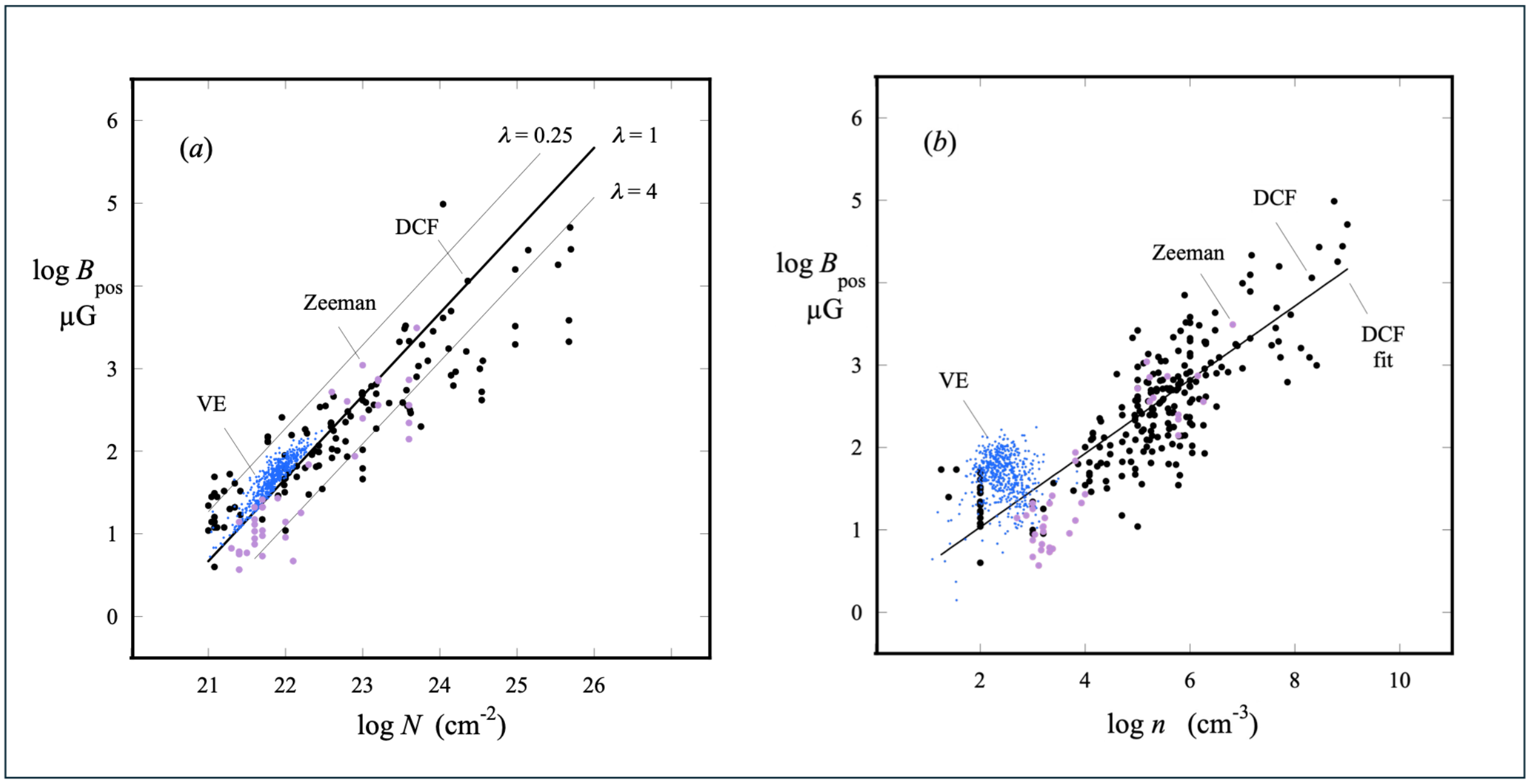

**Figure 9**. Magnetic field strength estimates $B_{\text{pos}}$ as functions of ($a$) column density $N$ and ($b$) number density $n$. Blue circles indicate GMCs in the GRS survey (RD10) based on the VE model of Section 3. Purple circles indicate estimates for clumps based on Zeeman observations (C12). Black circles indicate estimates for cores based on DCF analysis of infrared polarization (L22). Figure 9($a$) shows the magnetically critical line where $\lambda = 1$ and fainter lines where $\lambda = 0.25$ and $\lambda = 4$. Figure 9($b$) shows a linear fit to the DCF data, with slope 0.45 ± 0.02, consistent with that in L22. The errors for clouds are $\sigma_{\log B_{\text{pos}}} \approx \sigma_{\log N} \approx 0.16$ based on the median $\delta M/M$ in RD10 Table 1. The errors for Zeeman clumps are $\sigma_{\log B_{\text{pos}}} \approx 0.2$ based on typical signal-to-noise ratio = 3 (MB21). The errors for DCF cores are $\sigma_{\log B_{\text{pos}}} \approx 0.3$ (L22).

The foregoing magnetic VE analysis suggests that the gas in observed GMCs has a significantly greater ratio of magnetic to gravitational energy than the gas in star-forming clumps or cores. The strong fields of a subcritical cloud can limit its star formation more than the weak fields of a supercritical cloud, as discussed in Section 7.1.3. This concept is well known from VE theory (Mestel & Spitzer 1956; Mouschovias & Spitzer 1966) and from simulations (Krumholz & Federrath 2019; KF19). However, until now it has not been tested with supercritical and subcritical data based on observations (P23).

The observed virial parameters of clouds in M31 have recently been analyzed by a novel method, which finds that most of the clouds analyzed cannot be in a state of global collapse (Krumholz et al. 2025; K25). This conclusion is consistent with the finding in Figure 9 that CO clouds in M31 are generally magnetically subcritical.

### 5.2. Magnetic field strength in relation to gas flows

The VE estimates of mass-to-flux ratio in Figure 8 indicate that the magnetic fields in many CO survey clouds are strong enough to inhibit their gravitational contraction. The same data indicate that the magnetic energy of cloud fields is greater than the kinetic energy of cloud gas flows: strong fields channel gas flow along the field direction, whereas weaker fields are entrained along the direction of the flow. This property is quantified by the Alfvén Mach number $M_A \equiv \sigma_v/v_A$. Here $v_A = B_{tot}/\sqrt{4\pi\rho}$ is the Alfvén speed and $\rho$ is the local gas density. When a strong field controls the direction of flow, $M_A < 1$ and the turbulence is "sub-Alfvénic." When weak field lines are entrained by the flow, $M_A > 1$ and the turbulence is "super-Alfvénic" ( KF19).

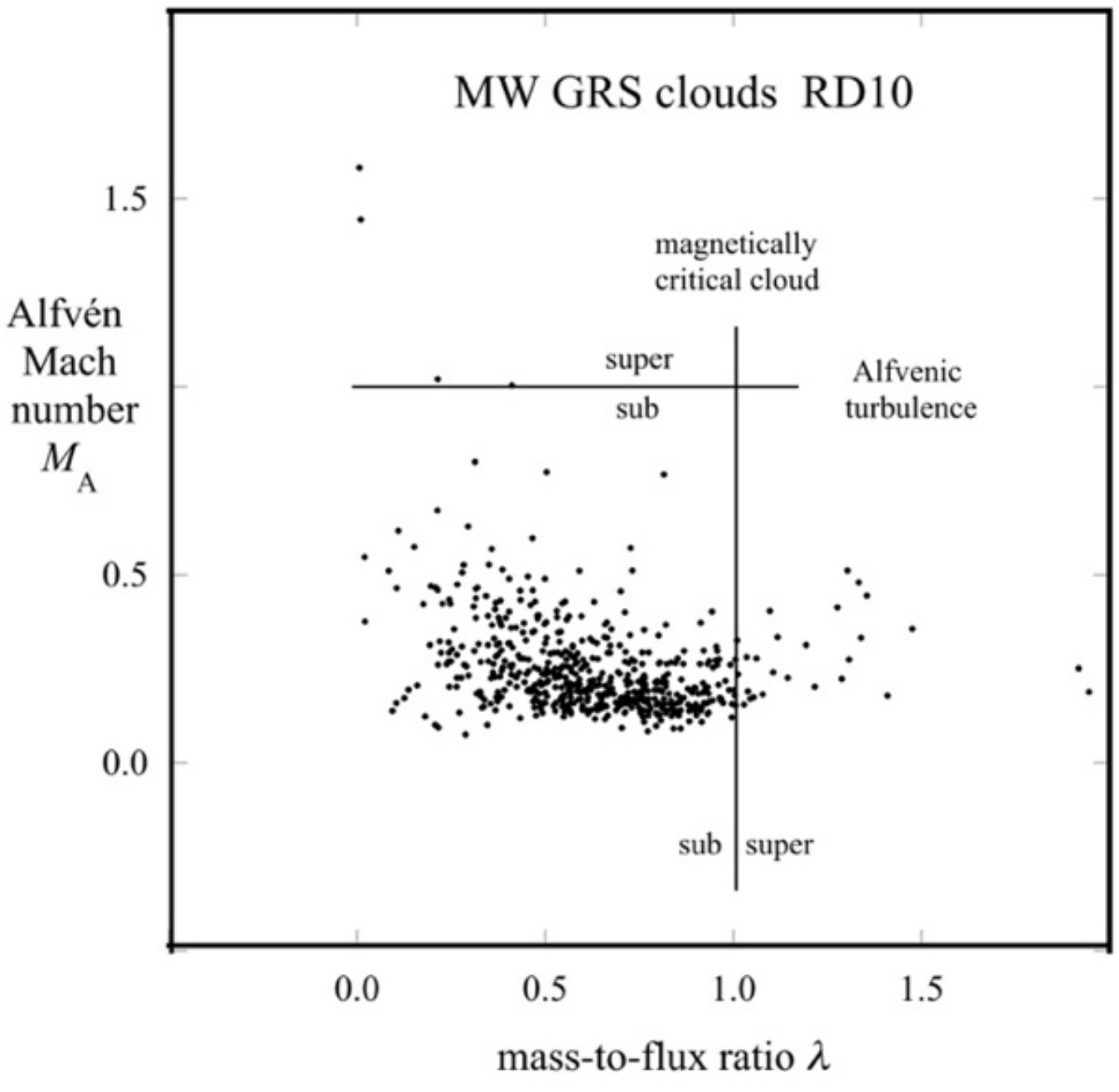

**Figure 10.** Alfvén Mach number $M_A$ plotted against mass-to-flux ratio $\lambda$ for GRS survey clouds. Their median values are $\lambda_{\mathrm{med}} = 0.61$ and $(M_A)_{\mathrm{med}} = 0.22$. These values indicate that the typical GRS cloud is magnetically subcritical and that its turbulence is sub-Alfvénic. Hence the magnetic field is strong enough to oppose global gravitational contraction, and it is also strong enough to channel turbulent flows and to resist field line entrainment.

The properties $M_A$ and $\lambda$ are correlated, since each quantity depends inversely on $B_{\mathrm{tot}}$. Their ratio can be written $M_A/\lambda = \sqrt{3}\pi c_\Phi \sigma_{\mathrm{v}}[R/(GM)]^{1/2}$ which is independent of $B_{\mathrm{tot}}$. Figure 10 shows $M_A$ plotted against $\lambda$, based on GRS survey data. The median value of $M_A/\lambda$ in the GRS survey is 0.37, hence a typical value of $M_A$ is $(M_A)_{\mathrm{med}} \approx (M_A/\lambda)_{\mathrm{med}}\lambda_{\mathrm{med}} = 0.25$. Thus the magnetic field is "strong" both in resisting gravitational contraction and in channeling gas flows.

## 6. Magnetic properties of Milky Way "bones"

MW "bones" are large, filamentary molecular clouds in the inner MW. Forty-five large-scale filaments observed by the *Herschel Space Observatory* were catalogued by Zucker et al. 2018 (Z18; see also Soler et al. 2021). Of these, 18 filaments of "bone" or *"Herschel"* type were classified as "bonelike." Their maps are approximately continuous and their spectral line observations indicate that they are velocity-coherent. They lie mostly parallel to the Galactic plane, and they have aspect ratio greater than ~20. They have typical mass ~ $10^4\ M_\odot$, column density ~ $10^{22}\ \mathrm{cm}^{-2}$, and projected length ~30 pc. Their star formation efficiency is similar to that of Solar neighborhood clouds (Zhang et al. 2019). The majority of star formation in the MW may occur in such filamentary structures (Pillsworth & Pudritz 2024).

The far infrared polarization of ten such bones has been imaged with the *SOFIA* HAWC+ polarimeter. These observations have been analyzed to estimate their magnetic properties (Stephens et al. 2025 (S25), Coudé et al. 2026 (C26)). This work extends earlier studies of filamentary clouds in the Solar neighborhood, which found a tendency toward perpendicular orientation with increasing column density (*Planck* collaboration 2016, Soler 2019, Oers et al. 2026).

This section estimates magnetic properties of all 18 of the bonelike clouds in Z18. These estimates are based on the virial equations of Section 3, modified to account for cloud elongation.

### 6.1. Virial estimates of magnetic properties of prolate spheroidal clouds

To analyze magnetic properties of elongated clouds we assume that they are prolate-spheroidal analogs of the spherical clouds analyzed in Section 3. We assume they are threaded by uniform magnetic fields perpendicular to their long axes, based on *Planck* results noted above, on observations of DR21 (Ching et al. 2022, Pillai et al. 2026), and on bone observations (S25, C26). We caution that this assumption is an approximation to a more complex picture.

For a uniform prolate spheroid with mass $M$, short radius $R$, aspect ratio $A$, external pressure $P_e$, velocity dispersion $\sigma_v$, and uniform magnetic field $B$ perpendicular to the long axis, the virial equation is modified from equation (1) or (2) to a form with explicit dependence on $A$,

$$3M\sigma_v^2 - 4\pi P_e A R^3 - \frac{3GM^2 f(A)}{5AR}\left(1 - \frac{1}{\lambda^2}\right) = 0 \qquad (26)$$

where the geometrical correction factor $f(A)$ is

$$f(A) = \frac{\ln\left(A + \sqrt{A^2 - 1}\right)}{\sqrt{A^2 - 1}} \qquad (27)$$

based on expressions in BM92. The mass-to-flux coefficient consistent with these equations is $c_\Phi = 0.13$ as in Section 3.1. Equations (26) and (27) indicate that the gravitational and magnetic energy densities decrease, each multiplied by the same factor $f(A)$, as the spheroid is stretched along its long axis, with fixed $R$, $M$, and $B$. When $A = 1$, $f(A) = 1$. When $A \gg 1, f(A) \approx A^{-1}\ln(2A)$. Note that for a uniform parallel field, $1/\lambda^2$ in equation (26) is replaced by $A^2/\lambda^2$ while $f(A)$ and $c_\Phi$ are unchanged.

In equation (26) the internal structure form factor $a$ is omitted since the spheroid is uniform with $a = 1$, as in equation (1) for the uniform sphere. The normalized mass-to-flux ratio is $\lambda = (M/\Phi)/(M/\Phi)_{cr}$ where the flux is $\Phi = \pi R^2 AB$ and the critical mass-to-flux ratio is $(M/\Phi)_{cr} = c_\Phi G^{-1/2}$.

In VE the normalized mass-to-flux ratio $\lambda = (M/\Phi)/(M/\Phi)_{cr}$ is obtained by solving equation (26) for $\lambda$, giving

$$\lambda = \left[1 - \frac{15MR\sigma_v^2 - 20\pi A R^4 P_e}{3GM^2 f(A)}\right]^{-1/2} \quad . \qquad (28)$$

In equation (28), the fraction inside the square bracket is the ratio of the net kinetic energy density to the gravitational energy density. When the net kinetic energy density is zero, equation (28) indicates $\lambda = 1$, hence the spheroid is in magnetically critical equilibrium where the magnetic energy balances the gravitational energy and the internal pressure balances the external pressure.

As the external pressure $P_e$ increases, $\lambda$ decreases toward more subcritical values, as in the spherical case analyzed in Section 3. Then the magnetic field tends to dominate the spheroid support. On the other hand, as the velocity dispersion $\sigma_v$ increases, $\lambda$ increases toward more supercritical values. Then the internal pressure tends to dominate the spheroid support, requiring less magnetic field strength for equilibrium.

The magnetic field strength $B_{tot}$ is obtained from the definitions of $\lambda$ and $\Phi$ as $B_{tot} = mNG^{1/2}/(\lambda c_\Phi)$, where $m$ is the mean particle mass and $N$ is the mean column density along the magnetic field direction. Substituting the value for $c_\Phi$ from Section 3.1 yields

$$B_{tot} = \frac{3\pi mNG^{1/2}}{\sqrt{5}\zeta\lambda} \quad . \qquad (29)$$

Equation (29) with $\zeta = 0.53$ matches equation (1) of Crutcher et al. (2004) within ~6%, after accounting for the difference between mean particle mass and mean $H_2$ mass.

### 6.2. Comparison of magnetic properties of MW clouds and bones

We evaluated equation (28) for $\lambda$ using values of $R$ from the bone FWHM in S25 Table 1, $A$ from Z18 Table 3, the high-density tracer $\sigma_v$ from S25 Table 6, and bone mass $M$ from C26 Table 2. We assumed that $P_e$ has the same inner-galaxy value as in Section 3, since the galactocentric radius $R_{gal}$ has median ± standard deviation 5.2 ± 0.8 kpc for bones (C26) and 5.9 ± 1.1 for GRS clouds (RD10). In evaluating equation (29) for $B_{tot}$, we used median column densities $N$ from S25 Table 3. Each of the eight bones in Z18 which were not observed in the *SOFIA* survey (S25, C26) was assumed to have the median $\sigma_v$ of those bones which are listed in S25. Each S25 estimate of $B_{tot}$ is based on the Davis-Chandrasekhar-Fermi (DCF) method of polarization analysis (P23).

The DCF method was applied to a sequence of "sliding boxes" along the spine of each observed bone. In each box, the normalized mass to flux ratio $\lambda$ and plane-of-the-sky field strength $B_{\text{pos}}$were evaluated by assuming that the critical mass to flux ratio coefficient is $c_\Phi = (2\pi)^{-1}$ (Nakano & Nakamura 1978). The plane-of-the-sky field strength $B_{\text{pos}}$ was converted to $B_{\text{tot}} = (4/\pi)B_{\text{pos}}$ using the standard statistical correction factor, assuming a uniform distribution of polarization angles over $4\pi$ steradians (Pillai et al. 2016).

The resulting values of $\lambda$ and of $B_{\text{tot}}$ are summarized in Table 1 for 583 GRS clouds according to the spherical VE model fits in Section 4 ("VE clouds"), for 18 bones according to the prolate spheroid VE models in this section ("VE bones"), and for 10 bones according to DCF analysis of *SOFIA* polarization observations ("DCF bones" (S25, C26)).

Table 1

Mass-to-Flux Ratio and Magnetic Field Strength in MW Clouds and Bones

| Property | VE clouds | VE bones | DCF bones |
|---|---|---|---|
| $\lambda$ | 0.5 - 0.8 | 0.5 - 1.0 | 0.4 - 0.8 |
| $B_{\text{tot}}$ (μG) | 30 - 100 | 70 - 150 | 40 - 130 |

In Table 1, $\lambda$ and $B_{\text{tot}}$ are each represented by the range extending from one standard deviation below the median to one standard deviation above the median. The table shows that the VE clouds, VE bones, and DCF bones have similar mass-to-flux ratios. They are each magnetically subcritical, varying by a factor of ~2, over $\lambda \approx 0.4 - 1.0$. The total field strength $B_{\text{tot}}$ also has substantial overlap among the three types of estimate. The greatest difference is between VE bones and VE clouds, where $B_{\text{tot}}(\text{VE bones}) \approx (1.5 - 2)B_{\text{tot}}(\text{VE clouds})$.

**6.3. Interpretation of magnetic properties of MW clouds and bones**

In Table 1, the similar estimates of $\lambda$ for VE clouds and bones indicate that MCs in the inner MW are consistently magnetically subcritical, despite their differences in density ($n \approx 2 \times 10^2 - 2 \times 10^3\ \text{cm}^{-3}$) and aspect ratio ($A \approx 1 - 30$). These common values of $\lambda$ may arise because GRS clouds and bones in the inner MW have relatively low internal kinetic pressure in an environment of relatively high external pressure. If so, the VE of clouds and bones in the inner MW may be dominated by magnetic support against external pressure, with internal kinetic pressure and self-gravity playing lesser roles.

The VE bones have field strength $B_{\text{tot}}$ greater than for VE clouds by a factor ~1.5-2, possibly because $B_{\text{tot}} \propto N/\lambda$ according to equation (29), and because the bones and clouds have a ratio of median column densities $N$(bones)/ $N$(clouds) ≈1.7 while their median values of $\lambda$ are nearly equal.

## 7. Discussion

This section describes limitations and uncertainties in Section 7.1, and implications in Section 7.2.

### 7.1 Limitations and uncertainties

#### 7.1.1. Origins of $s > 2$ and $s = 2$.

This section clarifies how well-determined values of $s = 2 - 3$ and $s = 2$ can arise. As noted in Section 1, some cloud samples show $s = 2$ with relatively high precision (local MW clouds, Lewis et al. 2022) and others show $s = 2 - 3$ with high precision (inner MW clouds, Figure 1($g$), RD10). Other samples have $s \approx 2$ with lower precision (massive star-forming clumps, Figure 1($f$), Traficante et al. 2018). Still others have high-precision values of $s$ ranging from ~2 to ~3 depending on the subregion considered (Figure 1($b$) -1($e$), MD17). Among these cases, when a sample has a well-determined value of $s$ in the range $s = 2 - 3$, its clouds are likely to be magnetically subcritical if they are in VE with sufficient external pressure, and if the range of their radii $R$ includes the scale length $R_{0,\text{sub}}$ (sections 3 and 4).

In contrast, a well-determined value $s = 2$ can arise in several ways. A sample of subcritical clouds in VE with $R \gg R_{0,\text{sub}}$ may follow $s = 2$ as in equation (16). Also, supercritical clouds may follow $s = 2$ when their external pressure is negligible compared to their internal pressure. Then if the observed line width-size ratio $\sigma_{\text{V}}^2/R$ is approximately constant over the sample, equation (2) indicates that the surface density $\Sigma$ is also approximately constant (Heyer et al. 2009). Then the property $\Sigma =$ constant implies $M \propto R^2$, i.e. $s = 2$. In a third way, a sample of supercritical clouds may follow $s = 2$ if the clouds are generally close to their maximum stable virial equilibrium (Chièze 1987, F11). Then $M \approx 2\left[5\pi P_{\text{e}}/\left(G\mu_{\text{sup}}\right)\right]^{1/2} R^2$, indicating $s = 2$ if the range of $\left(P_{\text{e}}/\mu_{\text{sup}}\right)^{1/2}$ is small compared to the range of $R^2$.

### 7.1.2. Should virial analysis of $M(R)$ include a turbulent cascade?

The interstellar media of galaxies show extensive evidence for supersonic turbulent motions (Hennebelle & Falgarone 2012). However it remains uncertain how a cloud environment injects and maintains the turbulence in a cloud. It is also unclear how the turbulent energy transfers from the large scales of injection to the small scales of star-forming cores. The power-law cascade scaling assumed by Cen (2021) resembles those due to eddies in Kolmogorov turbulence. However a detailed simulation of an isothermal cascade finds that 2D velocity power spectra more closely resemble Burgers turbulence driven by shocks (Fensch et al. 2023). Also, the anisotropy and intermittency of MHD turbulence poses challenges for simulations and analytic representations (Hennebelle & Inutsuka 2019; HI19). Thus a self-consistent combination of turbulent cascade and virial equilibrium, including magnetic fields and environmental pressure, remains an important problem for future investigation.

### 7.1.3. How does star formation depend on the initial mass-to-flux ratio?

In many discussions of VE, star-forming evolution of a cloud depends in a simple way on $\lambda_{\mathrm{i}}$, the initial value of its mass-to-flux ratio: the magnetic field prevents global cloud collapse if $\lambda_{\mathrm{i}} < 1$ or it allows global collapse if $\lambda_{\mathrm{i}} > 1$. However, MHD simulations indicate that "global collapse" does not accurately describe star formation in MCs (Grudič et al. 2019; G19), and analysis of CO surveys reaches a similar conclusion in clouds of the MW (Paper 1) and in M31 (K25).

Instead, recent MHD simulations portray star-forming cloud evolution as a competition between multi-scale turbulent contraction into filamentary star-forming gas, and gas dispersal due to feedback from young massive stars. Feedback dispersal typically limits the star forming efficiency to $SFE \lesssim 10\%$ (G19). As a cloud evolves, the magnetic field resists flows which cross field directions, and it slows down the formation of dense filaments, clumps and cores. In a set of simulations where the initial virial parameter is $\alpha_{\mathrm{vir}} = 2$, the $SFE$ at the onset of dispersal is much lower for a strong initial field than for a weak initial field. For $\lambda_{\mathrm{i}} = 0.5$, $SFE = 0.5\%$ while for $\lambda_{\mathrm{i}} = 8$, $SFE = 4\%$ (Kim et al. 2021; K21).

Clouds with weak initial fields form more subcritical and supercritical condensations than clouds with strong initial fields. Most of these condensations are dispersed by feedback before they can form stars (Herrington et al. 2024; see also Offner et al. 2025). In an analysis of STARFORGE simulations (Guszejnov et al. (2022)), the gas which forms a low-mass protostar waits longer to start its protostellar accretion in a strong-field cloud than in a weak-field cloud, i.e. it waits 9.6 Myr (longer time) when $\lambda_{\mathrm{i}}$ = 0.42 (stronger field) than 4.5 Myr (shorter time) when $\lambda_{\mathrm{i}}$ = 4.2 (weaker field) (Kaalva et al. 2026).

In summary, a low initial MFR inhibits star formation according to VE analysis and MHD simulations. However, simulations provide a more quantitative description of multiscale collapse, cloud dispersal, and the effects of the magnetic field on the cloud $SFE$ as the cloud evolves.

#### 7.1.4. Star formation efficiency from ideal-MHD simulations

The ideal-MHD simulations based on the STARFORGE framework (G21) or on the radiation-MHD simulations of K21 form a sink cell in a region of dense gas by adding mass along field lines, likely due to turbulent compression. This process does not involve flux loss due to ambipolar diffusion, which is expected to occur on a much smaller disk scale. Many authors suggest that the $SFE$ of a GMC is set mainly by the larger-scale processes of turbulent shock compression and stellar feedback from massive stars (Chevance et al. 2020). In this case, high-resolution studies such as K21 may offer the best available MHD estimates of $SFE$, uncertain by a factor ~2 when initial turbulence and $\lambda_{\mathrm{i}}$ are specified.

### 7.2. Implications

The main implications of this work are as follows.

1. The mass-radius plots in Section 2 show that the property $M \propto R^s$ with $s = 2 - 3$ is significantly different from the Larson relation $M \propto R^2$. The property $s = 2 - 3$ applies to thousands of CO survey clouds (often called giant molecular clouds (GMCs)) in the MW and in nearby galaxies, including the LMC, M31, and M33.

2. The property $s = 2 - 3$ is consistent with the equations of virial equilibrium (VE), but only if the VE equations include contributions from the cloud net magnetic field $B_{\mathrm{tot}}$ and from sufficient values of its external pressure $P_{\mathrm{e}}$. In this case the cloud must be magnetically subcritical, with normalized mass-to-flux ratio $\lambda < 1$. Nonmagnetic, magnetically supercritical, and magnetically critical forms of VE do not fit the observed $M - R$ relations.

3. Magnetically subcritical VE is consistent with the CO survey observations of 583 clouds in the MW Galactic Ring Survey (RD10) according to solutions of the VE equations, and according to a subcritical VE model fit to the observed relation $M \propto R^s$, $s = 2.35 \pm 0.02$, with correlation coefficient 0.97 (Figure 6). Such subcritical VE model fits yield distributions of $P_\mathrm{e}$, $\lambda$, and field strength $B_\mathrm{tot}$ for ~1500 GMCs in the MW, M31 and M33 for the first time (Figure 8).

4. MW "bones" have values of mass $M$ and column density $N$ similar to those of MW GRS clouds. They also have subcritical values of $\lambda$, which are similar to those of MW GRS clouds, according to VE analysis and to DCF analysis of their far infrared polarization. Thus bones resemble MW GRS clouds in their subcritical magnetic fields despite their differences in elongation, perhaps because they have similar mass, column density, and external pressure.

5. The new evidence in this paper supports ideas originating from the early history of MC studies. Subcritical clouds were first proposed to explain how molecular clouds can be supported against global collapse (Mestel & Spitzer 1956, Mouschovias & Spitzer 1966, Shu, Adams & Lizano 1987). Early simulations of turbulent, magnetically subcritical clouds formed quiescent cores in agreement with observed values of $SFE$ (Li & Nakamura 2004, Nakamura & Li 2005). However, magnetic field estimates based on the Zeeman effect indicated that most Zeeman sources are magnetically supercritical (C12). This result led to models of GMCs which are "born supercritical" (KF19, Chevance et al. 2020).

6. On the other hand, it is now clear that the typical GMC has far greater mass and size, and far less density than the typical Zeeman source or the typical DCF core (Section 5.1). Thus the supercritical property of Zeeman and DCF sources is not necessarily a global property of GMCs, as assumed by some authors. Instead, GMCs may be globally subcritical while they harbor clumps and cores which are locally supercritical, resembling observed Zeeman and DCF sources.

7. This picture of subcritical GMCs with supercritical cores is supported by MHD simulations which indicate that GMCs with strong initial fields form filaments, clumps and cores more slowly than do GMCs with weak initial fields (K21). Fewer of these clumps and cores survive dispersal by feedback. At the onset of dispersal, the *SFE* for a strong initial field is ~0.5 %, while the *SFE* for a weak initial field is ~4%. The few surviving clumps and cores may nonetheless form stars efficiently, because their local fields are weak according to available Zeeman and DCF observations (Section 5.2).

8. The VE estimates of the mass-to-flux ratio (MFR) in GMCs in Section 4 are the first systematic study of GMC field strengths based on observations. Relatively few estimates of GMC

MFRs based on polarization are available in the MW due to confusion along the line of sight, and in external galaxies due to poor resolution and sensitivity (P23).

9. Among MHD simulations, those by K21 are the most detailed study of $SFE$ as a function of initial MFR. Other recent simulations also find a reduction in $SFE$ with decreasing $\lambda$ (Hix et al. 2023, Ibáñez-Mejia et al. 2022), and this general property is described in recent reviews (Chevance et al. 2020, KF19, HI19).

10. Many authors have discussed the proposal that the low $SFE$ in MCs arises because only the densest gas in MCs forms stars (Lada et al. 2012, Evans et al. 2014; Mattern et al. 2024) and that much of this densest gas is in dense cores (Benson & Myers 1989; Hacar et al. 2023; Paper 1). Paper 1 discussed $SFE$ in terms of virial parameter, but not in terms of field strength. This paper extends the approach of Paper I. We apply observation-based estimates of field strength to the problem of low $SFE$, following earlier simulations by Padoan et al. (2014), Li et al. (2015), and K21.

**8. Conclusions**

The main conclusions of this work are based on the results in Sections 1-6 and on the implications in Section 7.

1. CO surveys of thousands of giant molecular clouds (GMCs) in the Milky Way, the LMC, and other galaxies indicate mass-radius scaling laws $M \propto R^s$ with exponent $s = 2 - 3$ (Figures 1 and 2). This property is distinctly different from the third Larson relation $M \propto R^2$. Its interpretation in this paper leads to an improved understanding of the role of magnetic fields in star formation.

2. GMC survey data are analyzed with the equations of VE for a uniform turbulent sphere, including self-gravity, internal kinetic energy, internal magnetic energy, and external pressure $P_e$. For values of $P_e$ typical of the inner MW the data are consistent with VE only if the typical GMC has normalized mass-to-flux ratio $\lambda < 1$, i.e. only if it is magnetically subcritical. In contrast, lower values of $P_e$ typical of the outer MW imply GMCs which are magnetically supercritical (Figures 3 and 7).

3. The slope and correlation coefficient of the "line width - size relation" for a sample of clouds in VE can vary significantly with $P_e$, from shallow slope and poor correlation in the inner MW to steep slope and good correlation in the outer MW (Figure 4).

4. Solutions of the magnetically subcritical VE equations can be written $M \propto R^s$ with exponent $s = 2 - 3$, consistent with the observed range of $M - R$ relations. A formula for $\log M\,(\log R)$ is applied to fit the $\log M\,(\log R)$ relation for 583 GMCs observed in the Galactic Ring Survey (RD10), with correlation coefficient 0.97 (Figure 6). This formula is also applied to survey data for GMCs in M31 and M33 (Figure 8).

5. Magnetically subcritical VE fits to $\log M\,(\log R)$ predict distributions of $\lambda$, $B_{\mathrm{tot}}$, and $P_{\mathrm{e}}/k$ for more than 1500 GMCs in the MW, M31, and M33 (Figure 8). Typical GMCs in each galaxy are consistent with the magnetically subcritical VE model provided they have sufficient external pressure. For each property the median value is greatest in the MW ($\lambda = 0.67$, $B_{\mathrm{tot}} = 63$ μG, $P_{\mathrm{e}}/k = 4.3 \times 10^5$ $\mathrm{cm}^{-3}$ K) and least in M33. From galaxy to galaxy these properties rank in the same order as the molecular cloud mass, and as the inverse of the virial parameter. Among these galaxies, GMCs in the MW have the greatest molecular cloud mass, the strongest fields, the greatest external pressure, and the lowest virial parameter.

6. Elongated MW "bones" having column densities, masses and external pressures similar to those of MW GMCs are also magnetically subcritical, with $\lambda \approx 0.5 - 0.1$, according to VE analysis which accounts for their elongation. These VE estimates of $\lambda$ also appear consistent with estimates based on DCF analysis of their far infrared polarization. Bones and GMCs may be magnetically subcritical because their internal kinetic pressures alone are too low to support them against self-gravity and the external pressure of the inner MW. If so, substantial net magnetic pressure is also needed to maintain their equilibrium.

7. While most MW GMCs are magnetically subcritical according to the foregoing VE analysis, most MC clumps and cores are magnetically supercritical, according to observations of the Zeeman effect in MC clumps and to DCF analysis of far infrared polarization in low-mass cores (Figure 9).

8. GMCs may have low *SFE*, because (1) observations and analysis in this paper indicate that many GMCs are consistent with pressure-bounded, magnetically subcritical VE, (2) observations indicate that Zeeman clumps and DCF cores are magnetically supercritical, and (3) MHD simulations indicate that initially subcritical GMCs tend to form fewer filaments, clumps and cores before they are dispersed by feedback.

**Acknowledgements**

We thank the referee for a helpful review. We thank Mark Heyer, Bruce Elmegreen, and Woorak Choi for helpful comments. P.C.M. acknowledges Irwin Shapiro and Terry Marshall for their support. I.W.S. and S.C. acknowledge financial support for this work provided by NASA through award #08_0186 issued by USRA. I.W.S. additionally acknowledges support from NASA through award #10_0505 issued by USRA.

**References**

Ballesteros-Paredes, J., Hartmann, L., Vázquez-Semadeni, E., et al. 2011, MNRAS, 411, 65

Ballesteros-Paredes, J., Román-Zúñiga, C., Salomé, Q. et al. 2019, MNRAS, 490, 2648 (BP19)

Benedettini, M., Traficante, A., Olmi, L. et al. 2021, A&A, 654, 144

Benson, P., & Myers, P. 1989, ApJS, 71, 89

Bergin, E., & Tafalla, M. 2007, ARA&A, 45, 339

Bertoldi, F., & McKee, C. 1992, ApJ,395, 140 (BM92)

Blitz, L., Rosolowsky, E. 2004, ApJL, 612, L129

Blitz, L., Rosolowsky, E. 2006, ApJ, 650, 933

Brand, J., Wouterlout, J. 1995, A&A, 303, 851

Cen, R. 2021, ApJ, 906, 4

Chen, B.-Q., Li, G.-X., Yuan, H.-B. et al. 2020, MNRAS, 493, 351

Chevance, M., Kruijssen, J., Vazquez-Semadeni, E. et al. 2020, SSRv, 216, 50

Chièze, J. 1987, A&A, 171, 225

Ching, T., Qiu, K., Li, D. et al. 2022, ApJ, 941, 122

Choi, W., Bureau, M., Liu, L. et al. 2024, MNRAS, 531, 4045

Coudé, S., Stephens, I., Myers, P. et al., ApJS, 282, 2 (C26)

Crutcher, R. 1999, ApJ, 520, 706 (C99)

Crutcher, R., Nutter, D., Ward-Thompson, D. et al. 2004, ApJ, 600, 279

Crutcher, R. 2012, ARA&A, 50, 29 (C12)

Dib, S., Kim, J., Vázquez-Semadeni et al. 2007, ApJ, 661, 262

Elmegreen, B. 1989, ApJ, 338, 178 (E89)

Elmegreen, B., & Falgarone, E. 1996, ApJ, 471, 816

Evans, N., Heiderman, A., & Vutisalchavakul, N. 2014, ApJ, 782, 114

Evans, N., Heyer, M., Miville-Deschenes, M. et al. 2021, ApJ, 920, 126

Falgarone, E., Troland, T., Crutcher, R. et al. 2008, A&A, 487, 287 (F08)

Fensch, J., Bournaud, F., Brucy, N. et al. 2023, A&A, 672, 193

Field, G., Blackman, E., & Keto, E. 2011, MNRAS, 416, 710 (F11)

Franco, J., Rodríguez-Puebla, A., Ballesteros-Paredes, J. et al. 2025, MNRAS, 543, 2507

Ganguly, S., Walch, S., Clarke, S. et al. 2024, MNRAS, 528, 36

Gratier, P., Braine, J., Rodriguez-Fernandez, N. et al. 2010, A&A, 522, 3

Grudič, M., Hopkins, P., Lee, E. et al. 2019, MNRAS, 488, 1501(G19)

Grudič, M., Guszejnov, D., Hopkins, P. et al. 2021, MNRAS, 506, 2199 (G21)

Guszejnov, D., Grudić, M. Y., Offner, S. S. R., et al. 2022, MNRAS, 515, 4929

Hacar, A., Clark, S., Heitsch, F. et al. 2023, ASPC, 534, 153

Hennebelle, P., & Falgarone, E. 2012, A&ARv, 20, 55

Hennebelle, P., & Inutsuka, S. 2019, FrASS, 6, 5

Herrington, N., Dobbs, C., & Bending, T. 2024, MNRAS, 532, 1701

Heyer, M., Carpenter, J., & Snell, R. 2001, ApJ, 551, 852 (H01)

Heyer, M., & Dame, T. 2015, ARA&A, 53, 583

Heyer, M., Krawczyk, C., Duval, J. et al. 2009, ApJ, 699, 1092

Hix, R., He, C., & Ricotti, M. 2023, MNRAS, 522, 6203

Ibáñez-Mejia, J., Mac Low, M., & Klessen R. 2022, ApJ, 925, 1961

Kaalva, A., Offner, S., Fillipova, N. et al. 2026, ApJ, 1002, 131

Kainulainen, J., Beuther, H., Banerjee, R. et al. 2011, A&A, 530, 64

Kim, J.-G., Ostriker, E., & Fillipova, N. 2021, ApJ, 911, 128 (K21)

Kritsuk, A., Lee, C., & Norman, M. 2013, MNRAS, 436, 3247

Krumholz, M., & Federrath, C. 2019, FrASS, 6, 7 (KF19)

Krumholz, M., Lada, C., & Forbrich, J. 2025, OJAp, 8, 91(K25)

Lada, C., Forbrich, J., Lombardi, M. et al. 2012, ApJ, 745, 190

Lada, C., Forbrich, J., Petipas, G. et al. 2024, ApJ, 966, 193

Larson, R. 1981, MNRAS, 194, 809 (L81)

Lewis, J., Lada, C., & Dame, T. 2022, ApJ, 931, 9

Li, Z.-Y., Nakamura, F. 2004, ApJ, 609, 83

Li, P.-S., McKee, C., & Klein, R. 2015, MNRAS, 452, 2500

Liu, J., Qiu, K., & Zhang, Q. 2022, ApJ, 925, 30 (L22)

Lombardi, M., Alves, J., & Lada, C. 2011, A&A, 535, 16

Ma, Y., Wang, H., Li, C. et al. 2021, ApJS, 254, 3

May, J., Alvarez, H., Bronfman, L. 1997, A&A, 327, 325

Mattern, M., André, P., Zavagno, A., et al. 2024, A&A, 688, 163

McKee, C., Zweibel, E., Goodman, A. et al. 1993, in Protostars and Planets III, ed. E. Levy et al. (Tucson, AZ: Univ. Arizona Press), 327 (M93)

Mestel, L., & Spitzer, L. 1956, MNRAS, 116, 503

Miville-Deschênes, M., Murray, N., & Lee, E. 2017, ApJ, 834, 57 (MD17)

Morii, K., Sanhueza, P., Nakamura, F. et al. 2023, ApJ, 950, 148

Mouschovias, T., & Spitzer, L. 1966, ApJ, 210, 326

Muraoka, K., Konishi, A., Tokuda, K. et al. 2023, ApJ, 953, 164

Myers, P., & Basu, S. 2021, ApJ, 917, 35 (MB21)

Myers, P., Heyer, M., Stephens, I. et al. 2025, ApJ, 991, 210 (Paper 1)

Nakano T., & Nakamura, T. 1978, PASJ, 30, 67

Nakamura, F., & Li, Z.-Y. 2025, ApJ, 631, 41

Nieten, C., Neininger, N., Guélin, M. et al. 2006, A&A, 453, 459

Oers, J., Ristorcelli, I., Ferrière, K. et al. 2026, A&A, 711, 97O

Offner, S., Taylor, J., & Grudič, M. 2025, ApJ, 982, 138

Ostriker, E., & Kim, C. 2022, ApJ, 936, 137 (OK22)

Padoan, P., Federrath, C., Chabrier, G., et al. 2014, in Protostars and Planets VI, ed. H. Beuther, R. Klessen, C. Dullemond, et al. (Tucson, AZ: Univ. Arizona Press), 77

Pattle, K., Fissel, L., Tahani, M. et al. 2023, ASPC, 534, 193

Peters, T., Banerjee, R., Klessen, R. et al. 2011, ApJ, 729, 72

Pillai, T., Kauffmann, J., Wiesemeyer, H. et al. 2016, A&A, 591, 19

Pillai, T., Kauffmann, J., Soler, J. et al. 2026, ApJ, 1004, 247

Pillsworth, R., & Pudritz, R. 2024, MNRAS, 528, 209

Planck Collaboration, Ade, P. A., Aghanim, N. et al. 2016, A&A, 586, A138

Rathborne, J., Johnson, A., Jackson, J. et al. 2009, ApJS, 182, 131

Roman-Duval, J., Jackson, J., Heyer, M. et al. 2010, ApJ, 723, 492 (RD10)

Schinnerer, E., & Leroy, A. 2024, ARA&A, 62, 369

Seta, A., & McClure-Griffiths, N. 2025, MNRAS, 539, 1024

Shu, F., Adams, F., & Lizano, S. 1987, ARA&A, 25, 23

Singh, A., Matzner, C., Friesen, R., et al. 2021, ApJ, 922, 87

Sodroski, T. 1991, ApJ, 366, 95

Soler, J. 2019, A&A 629, A96

Soler, J., Beuther, H., Syed, J. et al. 2021, A&A, 651, L4

Solomon, P., Rivolo, A., Barrett, J. et al. 1987, ApJ, 319, 730 (S87)

Spitzer, L. 1978, Physical Processes in the Interstellar Medium (New York: Wiley), 239

Stephens, I., Coudé, S., Myers, P. et al. 2025, ApJ, 995, 181(S25)

Strittmatter, P. 1966, MNRAS, 132, 359

Sun, J., Leroy, A., Schinnerer, E. et al. 2020, ApJ, 901, 8

Tomisaka, K., Ikeuchi, S., & Nakamura, T. 1988, ApJ, 326, 28

Traficante, A., Duarte-Cabral, A., Elia, D. et al. 2018, MNRAS 477, 4975

Troland, T., & Crutcher, R. 2008, ApJ, 680, 457 (TC08)

Wienen, M., Wyrowski, F., Schuller, F. et al. 2012, A&A, 544, 146

Wong, T., Hughes, A., Tokuda, K. et al. 2019, ApJ, 885, 50

Zhang, M., Kainulainen, J., Mattern, M. et al. 2019, A&A, 622, 52

Zucker, C., Battersby, C., & Goodman, A., ApJ, 864, 153 (Z18)